\documentclass[12pt]{iopart}
\jl{1}
\eqnobysec

\usepackage{subeqn}

\newcommand{\LMP}{{\em Lett. Math. Phys.} }
\newcommand{\PTP}{{\em Prog. Theor. Phys.} }

\newcommand{\SPJ}{{\em Sov. Phys.--JETP} }
\newcommand{\CMP}{{\em Commun. Math. Phys.} }

\def\hf{\frac{1}{2}}
\def\beq{\begin{equation}} \def\eeq{\end{equation}}
\def\bseq{\begin{subequations}} \def\eseq{\end{subequations}}
\def\bea{\begin{eqnarray}} \def\eea{\end{eqnarray}}
\def\bsea{\begin{subeqnarray}} \def\esea{\end{subeqnarray}}

\let\ti=\tilde  

\newcommand{\dsum}[2]{\sum_{\stackrel{\scriptstyle #1}{#2}}}

\def\eql{\eqalign} 

\def\num{\numparts} 
\def\enum{\endnumparts} 

\let\nn=\nonumber
\def\beann{\begin{eqnarray*}} \def\eeann{\end{eqnarray*}}

   \let\de=\delta
 \let\z=\zeta

\newcommand{\eins}{\mbox{1 \hspace{-10.8pt} I}}

\def\0{\over } \def\1{\vec }     \def\2{{1\over2}} \def\4{{1\over4}}
\def\5{\bar }  \def\6{\partial } \def\7#1{{#1}\llap{/}}

\def\<{\langle } \def\>{\rangle }

\def\i{{\rm i}} \def\tr{\mbox{tr}}

\def\d{{\rm d}}
\def\tz{{\tt z}}
\def\e{{\rm e}}

 \def\sech{\mbox{\,sech}}

\begin{document}
\jl{1}

\title{Integrable semi-discretization of the coupled 
nonlinear Schr\"{o}dinger equations}

\author{Takayuki Tsuchida\footnote[1]{E-mail address: {\tt
      tsuchida@monet.phys.s.u-tokyo.ac.jp}}, Hideaki Ujino and Miki Wadati 
}

\address{Department of Physics, Graduate School of Science, \\
University of Tokyo, \\
Hongo 7--3--1, Bunkyo-ku, Tokyo 113--0033, Japan}

\address{}

\address{Received 29 April 1998, in final form 25 November 1998}

\begin{abstract}
A system of semi-discrete coupled nonlinear Schr\"{o}dinger equations is 
studied. To show the complete integrability of the model with 
multiple components, we extend the discrete version of the inverse 
scattering method for the single-component discrete nonlinear Schr\"{o}dinger 
equation proposed by Ablowitz and Ladik. By means of the extension, 
the initial-value problem of the model is solved. Further, the 
integrals of motion and the soliton solutions are constructed 
within the framework of the extension of the inverse scattering 
method.
\end{abstract}


\maketitle

\section{Introduction}

There has been a surge of interest in the family of nonlinear 
Schr\"{o}dinger (NLS) equations because of its many applications 
to various kinds of physical phenomena. In the remarkable 
papers \cite{ZS1,ZS2} Zakharov and Shabat solved the NLS model,
\beq
\eql{
\i \frac{\partial q}{\partial t}
 + \frac{\partial^2 q}{\partial x^2}
- 2 qrq =0
\\
\i \frac{\partial r}{\partial t}
 - \frac{\partial^2 r}{\partial x^2}
+ 2 rqr =0
}
\label{NLS}
\eeq
with $r= \mp q^\ast$ by means of the inverse scattering 
method (ISM). After another success of the ISM for the modified 
KdV equation \cite{Wadati0}, 
Ablowitz, Kaup, Newell and Segur \cite{AKNS} 
unified a class of soliton equations by employing
various time dependences of the scattering problem. Their formulation
 is called the AKNS formulation. 

A number of authors have studied extensions of the AKNS formulation 
and presented many models, which are integrable by the
ISM \cite{AS,AC}. Among such models, a system of coupled nonlinear 
Schr\"{o}dinger (CNLS) equations
\beq
\eql{
\i \frac{\partial q_j}{\partial t}
 + \frac{\partial^2 q_j}{\partial x^2}
- 2\sum_{k=1}^m q_k r_k \cdot q_j =0
\\
\i \frac{\partial r_j}{\partial t}
 - \frac{\partial^2 r_j}{\partial x^2}
+ 2\sum_{k=1}^m r_k q_k \cdot r_j =0
}
\hspace{10mm}
j=1,2, \ldots, m
\label{CNLS}
\eeq
is particularly remarkable in describing diverse physical 
phenomena
\cite{Berk,Manakov,Shulman,Ohta2,RL,Menyuk,Kivshar,Yang,Inoue,Hisakado1,Hisakado2,Karlsson,RLH}.
Manakov 
\cite{Manakov} considered the two-component CNLS equations 
(\eref{CNLS} with $m=2$) with $r_j=-q_j^\ast \; (j=1,2)$ as a model 
for propagation of two
polarized electromagnetic waves and applied the ISM to the model for
the first time. Interest has recently focused on the two-component 
CNLS equations in
studying explicit solutions \cite{Ohta2,RL}, 
the stability of solitary
waves \cite{Yang} and interactions between solitons in
birefringent optical fibers \cite{Karlsson,RLH} from a physical point of view. 

Very recently, the authors proposed a new extension of the ISM and
solved the coupled modified KdV (cmKdV) equations \cite{Tsuchida1}
\beq
\frac{\6 u_i}{\6 t} + 6 \Bigl( \sum_{j,k=0}^{M-1} C_{jk} u_j u_k 
\Bigr) \frac{\6 u_i}{\6 x} + \frac{\6^3 u_i}{\6 x^3} = 0 
\hspace{10mm} i = 0,1, \ldots, M-1
\label{cmKdV1}
\eeq
in the self-focusing case, which had been investigated by alternative 
approaches \cite{Svinolupov,Iwao}. 
By a transformation of variables, this
model is cast into a new coupled version of the Hirota 
equation \cite{Hirota3}, which describes wave propagation in optical
fibers, including higher-order effects.

On the other hand, some discrete versions of the ISM have been constructed 
and applied to some discrete
models \cite{Case,Flaschka,Manakov2,AL1,AL2,Levi}. Among those models,
the semi-discrete nonlinear Schr\"{o}dinger (sd-NLS) equation found
 by Ablowitz and Ladik \cite{AL2} 
\beq
\eql{
\i \frac{\partial q_n}{\partial t}
 + (q_{n+1}+q_{n-1}-2q_{n})
- q_n r_n  (q_{n+1}+q_{n-1}) =0
\\
\i \frac{\partial r_n}{\partial t}
 - (r_{n+1}+r_{n-1}-2r_{n})
+ r_n q_n  (r_{n+1}+r_{n-1}) =0
}
\label{AL}
\eeq
has been studied extensively because of its simplicity and
physical significance. They solved \eref{AL} under the rapidly 
decreasing boundary conditions, $q_n, r_n \to 0$ at $ n \to \pm
\infty$. 
The model 
\eref{AL} was also solved under other integrable boundary
conditions \cite{Common,Veks,Ahmad}.

The sd-NLS equation has attracted researchers 
to studies on various subjects, 
such as nonlinear lattices in condensed matter 
physics \cite{FD}, phase plane patterns \cite{Ross},
 breather solutions \cite{Bishop}, 
B\"{a}cklund transformations \cite{Li}, 
numerical experiments and homoclinic structure \cite{AH,Umeki1,Umeki2}, 
the dynamics of a discrete curve \cite{Doliwa} and surfaces
\cite{Hisakado5}, Hamiltonian structure and
the classical 
$r$-matrix representation \cite{Kulish,FT,Suris} and the quantization of 
the model \cite{Kulish}.

In an analogous way to the continuous theory, 
it is natural to consider a
generalization of the sd-NLS equation \eref{AL} with multiple
components, namely, 
\beq
\fl
\eql{
\i \frac{\partial q^{(j)}_n}{\partial t}
 + (q^{(j)}_{n+1}+q^{(j)}_{n-1}-2q^{(j)}_{n})
- \sum_{k=1}^m q^{(k)}_n r^{(k)}_n \cdot (q^{(j)}_{n+1}+q^{(j)}_{n-1}) =0
\\
\i \frac{\partial r^{(j)}_n}{\partial t}
 - (r^{(j)}_{n+1}+r^{(j)}_{n-1}-2r^{(j)}_{n})
+ \sum_{k=1}^m r^{(k)}_n q^{(k)}_n \cdot (r^{(j)}_{n+1}+r^{(j)}_{n-1}) =0
}
\hspace{5mm}
j=1, 2, \ldots, m .
\label{CNLS1}
\eeq
We call this model the semi-discrete coupled NLS (sd-CNLS)
equations. The model is expected to be important in various
applications, e.g. numerical simulations of the CNLS equations \eref{CNLS}. 
Hisakado \cite{Hisakado4} showed that the system \eref{CNLS1} 
is connected with the two-dimensional Toda lattice. An $N$-soliton solution
was obtained by Ohta \cite{Ohta}. It is noted that another 
scheme of integrable semi-discretization of the CNLS equations was reported by 
Merola, Ragnisco and Gui-Zhang \cite{Merola}.

In \cite{Tsuchida2}, the authors proposed a new extension of the 
discrete version of the ISM by Ablowitz and Ladik. Applying 
the extension, they solved the initial-value problem of 
the semi-discrete coupled modified KdV (sd-cmKdV) equations, or
 the coupled modified Volterra equations \cite{Hisakado4},
\beq
\fl
\hspace{15mm}
  \frac{\6 u^{(i)}_n}{\6 t} = \Bigl(1+ \sum_{j,k=0}^{M-1} C_{jk}
 u^{(j)}_n u^{(k)}_n \Bigr) (u^{(i)}_{n+1}-u^{(i)}_{n-1}) 
  \hspace{5mm} i=0, 1, \ldots, M-1 
 \label{cmV1}
\eeq
under some appropriate conditions. A systematic procedure for constructing 
conservation laws and multi-soliton solutions was also
given \cite{Tsuchida2}. Related results are obtained by means of 
Hirota's method \cite{Ohta,Hirota2}. 

In the present paper, we use a transformation of 
variables
\beq
\eql{
\i^n \e^{2\i t}q_n^{(j)} = v_n^{(2j-2)} + \i v_n^{(2j-1)}
\\
(-\i)^n \e^{-2\i t}r_n^{(j)} = -v_n^{(2j-2)} + \i v_n^{(2j-1)}
}
\hspace{5mm} j=1,2, \ldots, m 
\label{trans}
\eeq
which cast the sd-CNLS equations \eref{CNLS1} into the sd-cmKdV
 equations \eref{cmV1} with $C_{jk}=\de_{j, k}$, 
$u_n^{(i)} \to v_n^{(i)}$ and $M=2m$. 
We pull back the transformation \eref{trans} to the level 
of the Lax representation and give an explicit Lax pair for the sd-CNLS
equations \eref{CNLS1} for the first time. Following the method of 
\cite{Tsuchida2}, we can solve the initial-value problem of the
sd-CNLS equations \eref{CNLS1} with $r_n^{(j)}=-q_n^{(j)\, \ast}$
under the rapidly decreasing boundary conditions, 
$q_n^{(j)} \to 0$ at $ n \to \pm \infty$. Explicit forms of
conserved quantities and the $N$-soliton solution are also given 
within the framework of the ISM. 

The paper consists of the following. In section 2, we introduce a Lax
pair for the semi-discrete matrix NLS equation. Considering a reduction 
to the sd-CNLS equations, we obtain the Lax formulation and
conservation laws for the sd-CNLS equations. In section 3, we perform
the ISM for the sd-CNLS equations with directing our attention to
 the transformation
\eref{trans}. The initial-value problem is solved and the $N$-soliton
solution is given. The last section is devoted to
discussions. 

The main idea of the paper is based on a matrix representation and
some properties of the 
Clifford algebra, whose elements are anti-commutative. The proof 
of relations used in the paper is given in \cite{Tsuchida2}. 

\section{Lax representation and conservation laws}

\subsection{Lax pair for the semi-discrete matrix NLS equation}
We begin with a set of auxiliary linear equations
  \beq
  \Psi_{n+1} = L_n \Psi_n ~~\Psi_{n,t} = M_n \Psi_n.
  \label{scattering_problem1}
  \eeq
Here $\Psi_n$ is a $2l$-component column vector, and $L_n$, $M_n$ are
 $2l \times 2l$ matrices. 
The compatibility condition of \eref{scattering_problem1} is 
  given by 
  \beq
  L_{n,t} +L_n M_n - M_{n+1}L_n = O.
  \label{Lax equation}
  \eeq
We call $L_n$ and $M_n$ the Lax pair and \eref{Lax equation}
 (a semi-discrete version of) the zero-curvature condition, 
or simply, the Lax equation. 
 Let us introduce the following form for the Lax pair:
  \beq
\fl
  L_n
  =
  \tz \left[
  \begin{array}{cc}
    F_1  &  O \\
    O  &  O \\
  \end{array}
  \right]
  +
  \left[
  \begin{array}{cc}
   O  &  F_1 Q_n \\
   F_2 R_n  &  O \\
  \end{array}
  \right]
 +
  \frac{1}{\tz} \left[
  \begin{array}{cc}
    O  &  O \\
    O  &  F_2 \\
  \end{array}
  \right]
  =
   \left[
  \begin{array}{cc}
   \tz F_1  &  F_1 Q_n \\
    F_2 R_n  &  \frac{1}{\tz} F_2 \\
  \end{array}
  \right]
  \label{U_form}
  \eeq
  \bea
\fl  
M_n &=&
   \tz^2
  \left[
  \begin{array}{cc}
  \i I &  O \\
   O  &  O \\
  \end{array}
  \right]
  + \tz
  \left[
  \begin{array}{cc}
   O  &  \i Q_n \\
   \i F_2 R_{n-1} F_1 &  O  \\
  \end{array}
  \right]
 +  \left[
  \begin{array}{cc}
   -\i Q_n F_2 R_{n-1} F_1 + \i H_1  & O \\
   O & \i R_n F_1 Q_{n-1} F_2 + \i H_2 \\
  \end{array}
 \right]
 \nn \\
\fl &&
  + \frac{1}{\tz}
  \left[
  \begin{array}{cc}
   O  &  -\i F_1 Q_{n-1} F_2 \\
   -\i R_{n} &  O  \\
  \end{array}
  \right]
  + \frac{1}{\tz^2}
  \left[
  \begin{array}{cc}
   O  &  O \\
   O  &  -\i I \\
  \end{array}
  \right]
 \nn \\
\fl  &=&
  \i 
  \left[
  \begin{array}{cc}
   \tz^2 I -Q_n F_2 R_{n-1}F_1 + H_1 &  \tz Q_n-\frac{1}{\tz}F_1Q_{n-1}F_2 \\
   \tz F_2 R_{n-1}F_1 - \frac{1}{\tz}R_n &  
   -\frac{1}{\tz^2}I + R_n F_1 Q_{n-1}F_2 + H_2  \\
  \end{array}
  \right]
\label{V_form}
\eea
where $\tz$ is the spectral parameter which is time-independent.
 $I$ is the $l \times l$ unit matrix, $Q_n$ and $R_n$ are 
$l \times l$ matrices. The constant matrices $F_1$, $F_2$, 
$H_1$ and $H_2$ are assumed to be Hermitian and satisfy
\beq
(F_1)^2 = (F_2)^2 = I
\hspace{5mm}
[F_1, H_1] = [F_2, H_2] = O.
\label{F_H}
\eeq
Here $[ \, \cdot \, , \cdot \, ]$ denotes the commutator. 
Substituting \eref{U_form} and \eref{V_form} into \eref{Lax equation},
 we obtain a set of matrix equations
 \beq
\fl
 \eql{
\i Q_{n,t} +F_1 ( Q_{n+1} + Q_{n-1}) F_2 +H_1 Q_n-Q_n H_2
-F_1Q_{n+1}F_2R_n Q_n - Q_n R_n F_1Q_{n-1}F_2 = O
\\
 \i R_{n,t} -F_2 (R_{n+1}+R_{n-1})F_1 -R_n H_1+H_2R_n
+F_2 R_{n+1} F_1 Q_n R_n + R_n Q_n F_2 R_{n-1}F_1 = O.
}
\hspace{5mm} 
\label{mNLS1}
 \eeq
We call this model the semi-discrete (sd-) 
matrix NLS equation. 
The integrable model \eref{mNLS1} with $F_1 = F_2 = I, H_1=-I, H_2=I$ 
was found by Ablowitz, Ohta and Trubatch \cite{Trubatch,Trubatch2}. 

\subsection{Conservation laws}
\label{Conservation laws}

In this subsection, we present a method to construct 
local conservation laws for the sd-matrix NLS equation \eref{mNLS1}, 
which is a discrete version of the method in the continuous
 theory \cite{Tsuchida1}. 
We start from an explicit expression of \eref{scattering_problem1},
\beq
\fl
\left[
\begin{array}{c}
 \Psi_{1\, n+1}  \\
 \Psi_{2\, n+1}  \\
\end{array}
\right]
=
\left[
\begin{array}{cc}
 F_{1\, n} & S_n \\
 T_n & F_{2\, n} \\
\end{array}
\right]
\left[
\begin{array}{c}
 \Psi_{1\, n}  \\
 \Psi_{2\, n}  \\
\end{array}
\right]
\hspace{5mm}
\left[
\begin{array}{c}
 \Psi_{1\, n}  \\
 \Psi_{2\, n}  \\
\end{array}
\right]_t
=
\left[
\begin{array}{cc}
 A_n & B_n \\
 C_n & D_n \\
\end{array}
\right]
\left[
\begin{array}{c}
 \Psi_{1\, n}  \\
 \Psi_{2\, n}  \\
\end{array}
\right]
\label{cons2}
\eeq
where all the entries in vectors and matrices are 
assumed to be $l \times l$ square matrices. 

Introducing an $l \times l$ square matrix $\Gamma_n$ by
\beq
\Gamma_n \equiv \Psi_{2\, n} \Psi_{1\, n}^{-1}
\label{Gamma}
\eeq
we can show the following relations from \eref{Lax equation}
 and \eref{cons2} \cite{Tsuchida2}:
\bea
\fl (S_n \Gamma_n + F_{1\, n})_t (S_n \Gamma_n + F_{1\, n})^{-1}
=& 
A_{n+1}-(S_n \Gamma_n + F_{1\, n})A_n(S_n \Gamma_n + F_{1\, n})^{-1}
\nn \\
\fl & +B_{n+1}\Gamma_{n+1}-(S_n \Gamma_n 
+ F_{1\, n})B_n \Gamma_n (S_n \Gamma_n + F_{1\, n})^{-1}
\hspace{15mm}
\label{cons10}
\eea
\beq
\Gamma_{n+1} = (T_n + F_{2\, n}\Gamma_n)(F_{1\, n}+S_n \Gamma_n)^{-1}.
\label{cons3}
\eeq
Taking the trace on both sides of \eref{cons10}, we obtain
\beq
\tr \{\log (S_n \Gamma_n + F_{1\, n})\}_t =
\tr (A_{n+1}+B_{n+1}\Gamma_{n+1})-\tr (A_{n}+B_{n}\Gamma_{n}).
\label{eq_ref1}
\eeq
Assuming the form of $L_n$ as \eref{U_form}, we have
\beq
F_{1\, n} = \tz F_1
\hspace{4.5mm}F_{2\, n} = \frac{1}{\tz} F_2
\hspace{4.5mm} S_n = F_1 Q_n \hspace{4.5mm}
        T_n = F_2 R_n .
\label{cons5}
\eeq
Then  
\eref{eq_ref1} and \eref{cons3} are transformed into
\beq
\fl 
\Bigl\{ \tr \, \log \Bigl(I + \frac{1}{\tz}Q_n \Gamma_n \Bigr) \Bigr\}_t 
=\tr (A_{n+1}+B_{n+1}\Gamma_{n+1})-\tr (A_{n}+B_{n}\Gamma_{n})
\label{cons6}
\eeq
\beq
\fl 
\tz Q_n \Gamma_n = Q_n F_2 R_{n-1} F_1 +\frac{1}{\tz}Q_n F_2 Q_{n-1}^{-1}
(Q_{n-1}\Gamma_{n-1})F_1 -(Q_n \Gamma_n)F_1 (Q_{n-1}\Gamma_{n-1})F_1 .
\label{cons7}
\eeq
Note that \eref{cons6} has the form of the local conservation law. 
This suggests that $\tr \{\log (I+Q_n\Gamma_n /\tz)\}$ is a generator of the 
conserved densities for \eref{mNLS1}. 
We substitute the expansion of $Q_n \Gamma_n$ with respect to $1/\tz$,
\beq
Q_n \Gamma_n^{(-)} = \sum_{j=1}^{\infty} \frac{1}{\tz^{2j-1}} f^{(j)}_n 
\label{cons8}
\eeq
into \eref{cons7}. Then we obtain a recursion formula for $f_n^{(j)}$,
\beq
\fl 
f^{(j)}_n = Q_n F_2 R_{n-1}F_1 \de_{j,1} + Q_n F_2 
Q_{n-1}^{-1}f^{(j-1)}_{n-1}F_1
 - \sum_{k=1}^{j-1} f^{(k)}_n F_1 f^{(j-k)}_{n-1} F_1
\hspace{6mm} j=1, 2, \ldots \, .
\label{f_formula}
\eeq
Equation \eref{f_formula} yields $f_n^{(j)}$, for instance, 
%
\[
f^{(1)}_n =  Q_{n} F_2 R_{n-1}F_1 
\]
\[
f^{(2)}_n = Q_n R_{n-2} - Q_n F_2 R_{n-1} Q_{n-1} F_2 R_{n-2}.
\]
%
%
We substitute \eref{cons8} into $\tr \{ \log (I+Q_n \Gamma_n^{(-)}/\tz)\}$ 
and expand it with respect to $1/\tz$,
\[
\fl 
\tr \Bigl\{ \log \Bigl( I + \frac{1}{\tz^2}f_n^{(1)}
+ \frac{1}{\tz^4}f_n^{(2)}+ \frac{1}{\tz^6}f_n^{(3)}
+ \cdots \Bigr) \Bigr\}
= \tr \Bigl\{
\frac{1}{\tz^2}f_n^{(1)} + \frac{1}{\tz^4} \Bigl[f_n^{(2)}
        -\hf (f_n^{(1)})^2 \Bigr] + \cdots
\Bigr\}.
\]
Thus, the first two conserved densities given by this expansion are
\num
\bea
\fl 
J_n^{(-1)} &= \tr \{f_n^{(1)}\} = \tr \{ Q_n F_2 R_{n-1}F_1 \}
\label{J^-1}
\\
\fl 
J_n^{(-2)} &= \tr \Bigl\{ f_n^{(2)} - \hf (f_n^{(1)})^2 \Bigr\}
\nn \\
\fl 
 &= \tr \Bigl\{ 
Q_n R_{n-2} - Q_n F_2 R_{n-1} Q_{n-1}F_2 R_{n-2} -
         \hf (Q_n F_2 R_{n-1}F_1)^2 \Bigr\}.
\hspace{15mm}
\label{J^-2}
\eea
\enum
Similarly, we expand $Q_n \Gamma_n$ with respect to $\tz$,
\beq
Q_n \Gamma_n^{(+)} = \sum_{j=1}^{\infty} \tz^{2j-1} g^{(j)}_n .
\label{posi_exp}
\eeq
Substitution of \eref{posi_exp} 
into \eref{cons7} yields a recursion formula for $g_n^{(j)}$,
\beq
\fl
g^{(j)}_n = -Q_n R_n \de_{j,1}+Q_n F_2 Q_{n+1}^{-1}g^{(j-1)}_{n+1}F_1
+Q_n F_2 Q_{n+1}^{-1} \sum_{k=1}^{j-1}g^{(k)}_{n+1} F_1 g^{(j-k)}_n 
\hspace{5mm} j=1,2, \ldots \, .
\label{g_rec}
\eeq
From formula \eref{g_rec}, 
the first three of the coefficients $g_n^{(j)}$ are given by
\[
\fl g^{(1)}_n = -Q_n R_n
\]
\[
\fl g^{(2)}_n = -Q_n F_2 R_{n+1} F_1 (I-Q_n R_n)
\]
\bea
\fl g^{(3)}_n =& -Q_n R_{n+2} (I-Q_n R_n)
+ (Q_n F_2 R_{n+1}F_1)^2 (I-Q_n R_n)
\nn \\
\fl & + Q_n R_{n+2}F_1 Q_{n+1}R_{n+1}F_1 (I-Q_n R_n) .
\nn
\eea
We substitute \eref{posi_exp} into $\tr \{ \log (I+Q_n
\Gamma_n^{(+)}/\tz)\}$ and expand it with respect to $\tz$,
\bea
\fl 
& \tr \{ \log ( I + g_n^{(1)} + \tz^2 g_n^{(2)}
+ \tz^4 g_n^{(3)} + \cdots ) \}
\nn \\
\fl 
=& \tr \Bigl\{
\log (I+g_n^{(1)}) + 
\tz^2 g_n^{(2)}(I+g_n^{(1)})^{-1}
 + \tz^4 \Bigl[ g_n^{(3)}(I+g_n^{(1)})^{-1}
-\hf \{g_n^{(2)}(I+g_n^{(1)})^{-1}\}^2 \Bigr] + \cdots
\Bigr\} .
\nn 
\eea
Thus, the first three conserved densities in this expansion are
\num
\bea
\fl 
J_n^{(0)} &= \tr \{\log (I + g_n^{(1)})\} = \tr \{\log (I-Q_n R_n)\}
\\
\fl 
J_n^{(1)} &= \tr \{ g_n^{(2)}(I+g_n^{(1)})^{-1}\}
         = \tr \{ -Q_n F_2 R_{n+1} F_1\}
\label{J^1}
\\
\fl 
 J_n^{(2)} &= \tr \Bigl[ g_n^{(3)}(I+g_n^{(1)})^{-1}
-\hf \{g_n^{(2)}(I+g_n^{(1)})^{-1}\}^2 \Bigr] 
\nn \\
\fl 
&= 
\tr \Bigl\{ -Q_n R_{n+2} + Q_n R_{n+2} F_1 Q_{n+1}R_{n+1}F_1 
 + \hf (Q_n F_2 R_{n+1}F_1)^2 \Bigr\}. \hspace{15mm}
\eea
\enum

The generator of the conserved densities, $\tr \{ \log (I+Q_n
\Gamma_n/\tz)\}$, is shown to be related with a time-independent
subset of scattering data defined later (see the appendix).

 \subsection{Reduction of the Lax pair and the conservation laws 
for the semi-discrete coupled NLS equations}
 \label{}
 In this subsection, we show a reduction of the sd-matrix
 NLS equations to the sd-CNLS equations. We recursively 
 define $2^{m-1} \times 2^{m-1}$ matrices $F_1^{(m)}$, $F_2^{(m)}$, 
$H_1^{(m)}$, $H_2^{(m)}$, $Q_n^{(m)}$ and $R_n^{(m)}$ by
 \beq
\fl
 F^{(1)}_1 = 1 \hspace{3mm} F^{(1)}_2 = 1\hspace{3mm}
 H_1^{(1)} = -1 \hspace{3mm} H_2^{(1)} = 1
 \label{rec1}
 \eeq
 \beq
\fl
 F^{(m+1)}_1
 = 
 \left[
 \begin{array}{cc}
  F^{(m)}_1 &   \\
    & - F^{(m)}_2 \\
 \end{array}
 \right]
\hspace{5mm}
 F^{(m+1)}_2
 = 
 \left[
 \begin{array}{cc}
  F^{(m)}_2 &   \\
    &  F^{(m)}_1 \\
 \end{array}
 \right]
 \label{rec2}
 \eeq
 \beq
\fl
 H^{(m+1)}_1
 = 
 \left[
 \begin{array}{cc}
  H^{(m)}_1 - I_{2^{m-1}} &  \\
    &    H^{(m)}_2 + I_{2^{m-1}} \\
 \end{array}
 \right]
\label{rec3}
\eeq
\beq
\fl
 H^{(m+1)}_2
 = 
 \left[
 \begin{array}{cc}
  H^{(m)}_2 - I_{2^{m-1}} &  \\
    &  H^{(m)}_1 + I_{2^{m-1}} \\
 \end{array}
 \right]
 \label{rec4}
\eeq
 \beq
\fl
 Q^{(1)}_n = q^{(1)}_n \hspace{5mm}
 R^{(1)}_n = r^{(1)}_n
 \label{QR_def1}
 \eeq
 \beq
\fl
 Q^{(m+1)}_n
 = 
 \left[
 \begin{array}{cc}
  Q^{(m)}_n &  q^{(m+1)}_n I_{2^{m-1}} \\
  r^{(m+1)}_n I_{2^{m-1}} &  -R^{(m)}_n \\
 \end{array}
 \right]
%
\hspace{5mm}
 R^{(m+1)}_n
 = 
 \left[
 \begin{array}{cc}
  R^{(m)}_n &  q^{(m+1)}_n I_{2^{m-1}} \\
  r^{(m+1)}_n I_{2^{m-1}} &  -Q^{(m)}_n \\
 \end{array}
 \right].
 \label{QR_def3}
 \eeq
Here $I_{2^{m-1}}$ is the $2^{m-1} \times 2^{m-1}$ unit matrix.
 It is readily seen that \eref{F_H} is satisfied. 
For 
the matrices defined by \eref{rec1}--\eref{QR_def3}, 
 we can prove the following relations:
 \beq
 Q^{(m)}_n R^{(m)}_n = R^{(m)}_n Q^{(m)}_n 
        = \sum_{j=1}^{m} q^{(j)}_n r^{(j)}_n \cdot I_{2^{m-1}}
 \label{Q_R_prod}
 \eeq
\[
 H^{(m)}_1 Q^{(m)}_n - Q^{(m)}_n H^{(m)}_2 = -2 F_1^{(m)}
         Q^{(m)}_n F_2^{(m)}
\]
\[
- R^{(m)}_n H^{(m)}_1 + H^{(m)}_2 R^{(m)}_n = 2 F_2^{(m)}
         R^{(m)}_n F_1^{(m)}
\]
by induction. Then substituting $Q^{(m)}_n$, $R^{(m)}_n$, etc. 
into $Q_n$, $R_n$, etc. in the sd-matrix NLS equation \eref{mNLS1}, we obtain 
the sd-CNLS equations 
\beq
\fl
\eql{
\i \frac{\partial q^{(j)}_n}{\partial t}
 + (q^{(j)}_{n+1}+q^{(j)}_{n-1}-2q^{(j)}_{n})
- \sum_{k=1}^m q^{(k)}_n r^{(k)}_n \cdot (q^{(j)}_{n+1}+q^{(j)}_{n-1}) =0
\\
\i \frac{\partial r^{(j)}_n}{\partial t}
 - (r^{(j)}_{n+1}+r^{(j)}_{n-1}-2r^{(j)}_{n})
+ \sum_{k=1}^m r^{(k)}_n q^{(k)}_n \cdot (r^{(j)}_{n+1}+r^{(j)}_{n-1}) =0
}
\hspace{5mm}
j=1, 2, \ldots, m .
\label{eq_ref44}
\eeq
For instance, the Lax matrix $L_n$ for the two-component sd-CNLS 
equations (\eref{eq_ref44} with $m=2$) is given by
\beq
L_n =
\left[
\begin{array}{cccc}
 \tz & 0 & q^{(1)}_n & q_n^{(2)}  \\
 0 & -\tz & -r_n^{(2)} & r_n^{(1)}  \\
 r_n^{(1)} & q_n^{(2)} & 1/\tz & 0 \\
 r_n^{(2)} & -q_n^{(1)} & 0 & 1/\tz \\
\end{array}
\right].
\label{L_n_example}
\eeq
%
%
%

In what follows, we set 
$r^{(j)}_n = - q^{(j) \ast}_n \; (j = 1, 2, \ldots, m)$ 
and consider a self-focusing case of the 
sd-CNLS equations,
\beq
\fl 
\i \frac{\partial q^{(j)}_n}{\partial t}
 + (q^{(j)}_{n+1}+q^{(j)}_{n-1}-2q^{(j)}_{n})
+ \sum_{k=1}^m |q^{(k)}_n|^2 \cdot (q^{(j)}_{n+1}+q^{(j)}_{n-1}) =0
\hspace{5mm} j = 1,2, \ldots, m.
\label{CNLS4}
\eeq
In this case, the relation \eref{Q_R_prod} for 
$Q_n^{(m)}$ and $R_n^{(m)}$ becomes
\beq
 Q^{(m)}_n R^{(m)}_n =  R^{(m)}_n Q^{(m)}_n = -\sum_{j=1}^{m} |q^{(j)}_n|^2
 \cdot I_{2^{m-1}}.
\label{eq_ref2}
\eeq
In addition, a simple relation between $Q_n^{(m)}$ and $R_n^{(m)}$ holds,
 \beq
 R_n^{(m)} = - Q_n^{(m)\, \dagger}
 \label{R_Q}
 \eeq
where the symbol $\dagger$ denotes the Hermitian conjugate. 
The relations \eref{eq_ref2} and \eref{R_Q} 
play an essential role in the ISM. 

Because the sd-CNLS equations are given as a reduction of the
sd-matrix NLS equation \eref{mNLS1}, the results in 
 section \ref{Conservation laws} assure the existence of an infinite number of conservation laws 
for the sd-CNLS equations. 
Explicit forms of the first four 
conserved densities for the sd-CNLS equations 
\eref{eq_ref44} are given by 
\beq
\fl
\log \Bigl( 1-\sum_j q^{(j)}_n r^{(j)}_n \Bigr)
\label{eq_ref3}
\eeq
\beq
\fl
\left\{
\begin{array}{l}
\displaystyle
\e^{4 \i t} (-1)^{n} (q^{(j)}_{n+1} q^{(k)}_n-q^{(j)}_{n} q^{(k)}_{n+1})
 \hspace{5mm} \forall \; j,\, k
\vspace{2mm}
\\
\displaystyle
\displaystyle
\e^{-4 \i t} (-1)^{n} (r^{(j)}_{n+1} r^{(k)}_n-r^{(j)}_{n} r^{(k)}_{n+1})
 \hspace{5mm} \forall \; j,\, k
\vspace{2mm}
\\
\displaystyle
q^{(j)}_{n+1}r^{(k)}_n + q^{(j)}_n r^{(k)}_{n+1}
 \hspace{5mm} \forall \; j,\, k
\vspace{2mm}
\\
\displaystyle
\sum_j q^{(j)}_{n+1} r^{(j)}_n \hspace{7mm}
\sum_j q^{(j)}_{n} r^{(j)}_{n+1}
\\
\end{array}
\right.
\label{conserved_q}
\eeq
\bea
\fl
\Bigl( 1 - \sum_{j} q^{(j)}_{n+1} r^{(j)}_{n+1} \Bigr)
\cdot \sum_{j} ( q^{(j)}_{n+2} r^{(j)}_n + q^{(j)}_{n} r^{(j)}_{n+2} )
-\frac{1}{2} \Bigl\{ \sum_{j} 
(q^{(j)}_{n+1} r^{(j)}_n - q^{(j)}_n r^{(j)}_{n+1}) \Bigr\}^2 
 - \sum_{j} q^{(j)}_{n+1} r^{(j)}_{n+1}
        \cdot \sum_{j} q^{(j)}_{n} r^{(j)}_{n}
\nn \\
\label{}
\eea
\bea
\fl
&&
 \Bigl( 1 - \sum_{j} q^{(j)}_{n+2} r^{(j)}_{n+2} \Bigr)
 \Bigl( 1 - \sum_{j} q^{(j)}_{n+1} r^{(j)}_{n+1} \Bigr)
 \cdot \sum_{j} (q^{(j)}_{n+3} r^{(j)}_n - q^{(j)}_{n} r^{(j)}_{n+3})
\nn \\
\fl 
&& 
- \Bigl(1- \sum_{j} q^{(j)}_{n+1} r^{(j)}_{n+1} \Bigr)
\cdot \sum_{j} (q^{(j)}_{n+2} r^{(j)}_n + q^{(j)}_{n} r^{(j)}_{n+2})
\cdot \Bigl\{
\sum_j ( q^{(j)}_{n+1} r^{(j)}_{n} - q^{(j)}_{n} r^{(j)}_{n+1})
 + \sum_{j} ( q^{(j)}_{n+2} r^{(j)}_{n+1} - q^{(j)}_{n+1} r^{(j)}_{n+2})
\Bigl\}
\nn \\
\fl 
&& 
- \Bigl(1 - \sum_{j} q^{(j)}_{n+1} r^{(j)}_{n+1} \Bigr)
\Bigl\{
\sum_j (q^{(j)}_{n+1} r^{(j)}_{n} - q^{(j)}_{n} r^{(j)}_{n+1})
\cdot \sum_j q^{(j)}_{n+2}r^{(j)}_{n+2}
+ \sum_j (q^{(j)}_{n+2} r^{(j)}_{n+1} - q^{(j)}_{n+1} r^{(j)}_{n+2})
\cdot \sum_j q^{(j)}_{n}r^{(j)}_{n}
\Bigr\}
\nn \\
\fl 
&&
+ \frac{1}{3} 
\Bigl\{
\sum_j (q^{(j)}_{n+1} r^{(j)}_{n} - q^{(j)}_{n} r^{(j)}_{n+1})
\Bigr\}^3
+ \sum_j (q^{(j)}_{n+1} r^{(j)}_{n} - q^{(j)}_{n} r^{(j)}_{n+1})
\cdot \sum_j q^{(j)}_{n+1} r^{(j)}_{n+1}
\cdot \sum_j q^{(j)}_{n}r^{(j)}_{n} .
\label{}
\eea
A straightforward calculation shows that all the entries in 
\eref{conserved_q} are conserved densities. 
It is noted that the lower two densities in
\eref{conserved_q} correspond to the conserved densities 
$q_j r_k$ and $\sum_j q_j r_j$ for the continuous CNLS
 equations \eref{CNLS}. 

It is also remarkable that higher than second conserved
densities for \eref{eq_ref44} have expressions such as
\[
\prod_{k=1}^{i-1} \Bigl(1 - \sum_j q_{n+k}^{(j)}r_{n+k}^{(j)} \Bigl)
\cdot \sum_j \{ q_{n+i}^{(j)}r_n^{(j)} + (-1)^i q_n^{(j)}r_{n+i}^{(j)} 
\} + \cdots \hspace{5mm} (i \ge 2)
\]
which split into two independent densities with simpler structures,
\[
\prod_{k=1}^{i-1} (1 - q_{n+k} r_{n+k} )
\cdot q_{n+i} r_n  + \cdots \hspace{5mm} (i \ge 2)
\]
and
\[
\prod_{k=1}^{i-1} (1 - q_{n+k} r_{n+k} )
\cdot q_n r_{n+i} + \cdots \hspace{5mm} (i \ge 2)
\]
in the single-component case (cf equation \eref{AL}). This fact implies that a 
direct recursion formula of the conserved densities for the sd-CNLS
equations might be, if it exists, very complicated. 
We have obtained the first four conserved densities concisely by use of the
recursion formula of the conserved densities for the sd-matrix NLS
equation \eref{mNLS1}.

\section{Inverse scattering method}
\label{ISM}

In this section we investigate the scattering and inverse scattering 
problems associated with 
the $2l \times 2l$ ($l=2^{m-1}$) matrix \eref{U_form},
\beq
\left[
\begin{array}{c}
 \Psi_{1 \, n+1} \\
 \Psi_{2 \, n+1} \\
\end{array}
\right]
=
\left[
\begin{array}{cc}
 \tz F_1 & F_1 Q_n \\
 F_2 R_n & \frac{1}{\tz}F_2 \\
\end{array}
\right]
\left[
\begin{array}{c}
 \Psi_{1 \, n}   \\
 \Psi_{2 \, n}   \\
\end{array}
\right] 
\label{scattering0}
\eeq
to solve the sd-CNLS equations \eref{CNLS4}. Here and hereafter 
the superscripts $(m)$ of $F_1^{(m)}$, $F_2^{(m)}$, $Q_n^{(m)}$ and 
$R_n^{(m)}$ are often omitted for convenience. 
To simplify the analysis, we consider a gauge transformation
\beq
\Phi_n = g_n \Psi_n
\hspace{5mm}
g_n = 
\left[
\begin{array}{cc}
 \e^{\i \frac{\pi}{4}n}(F_1)^n \e^{-\i H_1 t} &  \\
      &  \e^{-\i \frac{\pi}{4}(n-1)} (F_2)^n \e^{-\i H_2 t}\\
\end{array}
\right] .
\label{gauge}
\eeq
Then the scattering problem \eref{scattering0} is changed into
\beq
\left[
\begin{array}{c}
 \Phi_{1 \, n+1} \\
 \Phi_{2 \, n+1} \\
\end{array}
\right]
=
\left[
\begin{array}{cc}
 z I & \ti{Q}_n \\
 \ti{R}_n & \frac{1}{z}I \\
\end{array}
\right]
\left[
\begin{array}{c}
 \Phi_{1 \, n}   \\
 \Phi_{2 \, n}   \\
\end{array}
\right]
\label{scattering_problem2}
\eeq
where the transformed spectral parameter and potentials are 
\beq
\fl
  z = \tz \e^{\i \frac{\pi}{4}} 
\hspace{5mm}
\tilde{Q}_n = \i^n (F_1)^n \e^{-\i H_1 t} Q_n \e^{\i H_2 t}(F_2)^n
\hspace{5mm} \tilde{R}_n = (-\i)^n (F_2)^n \e^{-\i H_2 t}
        R_n \e^{\i H_1 t}(F_1)^n .
\label{QR_trans}
\eeq
%

The constraints \eref{R_Q} and \eref{eq_ref2} lead to those for 
$\ti{Q}_n$ and $\ti{R}_n$,
\beq
 \ti{R}_n=-\ti{Q}_n^{\, \dagger}
 \hspace{5mm} \ti{Q}_n \ti{R}_n = \ti{R}_n \ti{Q}_n =
         -\sum_{j=1}^{m} |q_n^{(j)}|^2 \cdot I \equiv 
 -\sigma_n I.
\label{eq_ref91}
\eeq
We assume the rapidly decreasing boundary conditions,
\beq
 \ti{Q}_n, \ti{R}_n \to O \hspace{5mm} {\rm as}
 \hspace{5mm} n \to \pm \infty.
 \label{boundary}
 \eeq
Considering the time dependence of the scattering data, we can solve the
initial-value problem of the sd-CNLS equations \eref{CNLS4}. 
Some of the main ideas in the following are an extension and 
a modification of the analyses for the matrix KdV 
equation \cite{Wadati3}, the matrix mKdV equations and the cmKdV
 equations \cite{Tsuchida1} (see \cite{Tsuchida2} for details). 

\subsection{Scattering problem}
Let $\Phi_n^{(1)} (z)$ and $\Phi_n^{(2)} (z)$ be 
matrix solutions of \eref{scattering_problem2} composed of
 $2l\,(=2^m)$ rows and $l\,(=2^{m-1})$ columns. 
We introduce the following matrix function of 
$\Phi^{(1)}$ and $\Phi^{(2)}$:
\beq
 W_n[\Phi^{(1)}, \Phi^{(2)}] 
\equiv \Phi_n^{(1)\, \dagger}\bigl(\frac{1}{z^{\ast}}\bigr)
 \Phi_n^{(2)}(z) .
 \label{}
 \eeq
This satisfies a recursion relation
\beq
W_{n+1}[\Phi^{(1)}, \Phi^{(2)}] = (I-\ti{Q}_n \ti{R}_n) W_n 
[\Phi^{(1)}, \Phi^{(2)}] = \rho_n W_n [\Phi^{(1)}, \Phi^{(2)}] 
\label{W_rec}
\eeq
where $\rho_n$ is defined by
\beq
\rho_n \equiv 1+ \sigma_n = 1+ \sum_{j=1}^{m} 
        |q^{(j)}_n|^2 .
\label{rho_def}
\eeq
%
%
%

 We introduce Jost functions $\phi_n$, $\5{\phi}_n$ and $\psi_n$,
 $\5{\psi}_n$ which satisfy the boundary conditions
 \bea
 \phi_n \sim 
 \left[
 \begin{array}{c}
   I  \\
   O  \\
 \end{array}
 \right]
 z^n
\hspace{5mm}
 \bar{\phi}_n \sim 
 \left[
 \begin{array}{c}
   O  \\
   -I  \\
 \end{array}
 \right]
 z^{-n}
 \hspace{7mm}
 {\rm as}~~~ n \rightarrow -\infty
 \label{phi_bar}
 \eea
and
 \bea
 \psi_n \sim 
 \left[
 \begin{array}{c}
   O  \\
   I  \\
 \end{array}
 \right]
 z^{-n}
\hspace{5mm}
 \bar{\psi}_n \sim 
 \left[
 \begin{array}{c}
   I  \\
   O  \\
 \end{array}
 \right]
 z^n
 \hspace{7mm}
 {\rm as}~~~ n \rightarrow +\infty.
 \label{psi_bar}
 \eea
Here $O$ and $I$ are, respectively, the $l \times l$ 
zero matrix and the $l \times l$ unit matrix. 
We can show that $\phi_n z^{-n}$, $\psi_n z^n$
 are analytic outside the unit circle ($|z|>1$) 
on the complex $z$ plane, and that $\5{\phi}_n z^n$, 
$\5{\psi}_n z^{-n}$ are analytic inside the unit circle ($|z|<1$) on
 the $z$ plane, when $\ti{Q}_n$ and $\ti{R}_n$ go to $O$
 sufficiently rapidly as $n \to \pm \infty$. 
 We assume the following summation representation of the Jost functions 
 $\psi_n$ and $\5{\psi}_n$:
 \beq
 \psi_n 
  =
\sum_{n'=n}^{\infty} z^{-n'} K (n,n')
\hspace{5mm}
 \bar{\psi}_n =
\sum_{n'=n}^{\infty} z^{n'} \bar{K} (n,n')
 \label{psi_form}
 \eeq
 where $K(n,n')$ and $\5{K}(n,n')$ are $z$-independent 
column vectors which consist of two $l \times l$ square matrices,
 \[
 K(n,n') =
 \left[
\begin{array}{c}
  K_1(n,n')  \\
  K_2(n,n')  \\
 \end{array}
 \right]
 \hspace{5mm}
 \bar{K}(n,n') =
 \left[
 \begin{array}{c}
  \bar{K}_1(n,n')  \\
  \bar{K}_2(n,n')  \\
 \end{array}
 \right].
 \]
We substitute \eref{psi_form} 
into \eref{scattering_problem2}. Equating the terms with the same 
power of $z$, we obtain
\bseq
\bea
K_1 (n,n) = O
\label{K1}
\\
K_2 (n,n) = \prod_{i=n}^\infty (I-\ti{R}_i \ti{Q}_i)^{-1} 
= \prod_{i=n}^\infty \rho_i^{-1} \cdot I 
\label{K3}
\\
\ti{Q}_n K_2 (n,n) = -K_1(n,n+1)
\label{K2}
\eea
\label{K_eq}
\eseq
and 
\bseq
\bea
\bar{K}_2 (n,n) = O
\label{K_bar1}
\\
\bar{K}_1 (n,n) = \prod_{i=n}^\infty (I-\ti{Q}_i \ti{R}_i)^{-1} 
= \prod_{i=n}^\infty \rho_i^{-1}  \cdot I 
\label{K_bar3}
\\
\ti{R}_n \bar{K}_1 (n,n) = -\bar{K}_2(n,n+1).
\label{K_bar2}
\eea
\label{K_bar_eq}
\eseq
Since a pair of the Jost functions $\phi_n$ and $\5{\phi}_n$, or 
 $\psi_n$ and $\5{\psi}_n$, forms a fundamental system of the 
solutions of the scattering problem \eref{scattering_problem2}, we can set
 %
 %
\bseq
 \bea
 \phi_n(z) = \bar{\psi}_n(z)A(z) + \psi_n(z)B(z)
 \label{phi_relation}
\\
 \bar{\phi}_n(z) = \bar{\psi}_n(z) \bar{B}(z) - \psi_n(z) \bar{A}(z).
 \label{phi_bar_relation}
\eea
\label{ref111}
\eseq
Here the coefficients $\{$$A(z)$, $\5{A}(z)$, $B(z)$, $\5{B}(z)$$\}$
 are $n$-independent $l \times l$ matrices which are called scattering data. 

To derive the formula of the ISM rigorously and 
concisely, we assume that $\ti{Q}_n$ and $\ti{R}_n$
 are on compact support. 
%
The result is, however, 
 valid for larger classes of the potentials $\ti{Q}_n$ and $\ti{R}_n$. 
Using the asymptotic behaviors of the Jost functions
 \eref{phi_bar}, \eref{psi_bar} and the relation \eref{W_rec}, we obtain
 \bea
 A(z) = W_\infty[ \5{\psi}, \phi]
\hspace{5mm}
 \5{A}(z) = - W_\infty[ \psi, \5{\phi}]
\label{rep2}
\eea
and 
\num
 \bea
\fl
\hspace{10mm}
 A^{\dagger}\bigl(\frac{1}{z^{\ast}}\bigr)A(z) +
B^{\dagger}\bigl(\frac{1}{z^{\ast}}\bigr)B(z) =
 \5{A}^{\dagger}\bigl(\frac{1}{z^{\ast}}\bigr)\5{A}(z) +
  \5{B}^{\dagger}\bigl(\frac{1}{z^{\ast}}\bigr)\5{B}(z) = 
\prod_{n=-\infty}^\infty \rho_n \cdot I
 \label{scat_rel1}
\\
 A^{\dagger}\bigl(\frac{1}{z^{\ast}}\bigr)\5{B}(z) =
 B^{\dagger}\bigl(\frac{1}{z^{\ast}}\bigr)\5{A}(z) .
\label{scat_rel3}
\eea
\label{A_B}
\enum
The expressions \eref{rep2} show
 that $A(z)$ and $\5{A}(z)$ are, respectively, analytic
 outside the unit circle ($|z|>1$) and 
inside the unit circle ($|z|<1$). 

 \subsection{Gel'fand-Levitan-Marchenko equations}
 \label{GLM_eq}
Multiplying $A(z)^{-1}$ and $\5{A}(z)^{-1}$ from the right to  
 \eref{phi_relation} and \eref{phi_bar_relation}, respectively, we have
\num
 \bea
 \phi_n(z) A(z)^{-1} = \bar{\psi}_n(z) + \psi_n(z)B(z)A(z)^{-1}
 \label{prep1}
\\
 \bar{\phi}_n(z) \bar{A}(z)^{-1} 
 = - \psi_n(z) + \bar{\psi}_n(z) \bar{B}(z) \bar{A}(z)^{-1}.
 \label{prep2}
\eea
\enum
We substitute \eref{psi_form} into the right-hand side of 
\eref{prep1} and operate on both sides 
 \[
 \frac{1}{2\pi \i }\oint_{C} \d z \; z^{-m-1} \hspace{5mm} (m \ge n)
 \]
where $C$ denotes a contour along the 
unit circle $|z|=1$. 
It should be noticed
 that $\phi_n z^{-n}$ and $A(z)$ are analytic outside the
unit circle $C$, $|z|>1$. The inverse of $A(z)$, i.e. $A(z)^{-1}$, is given by
 \[
 A(z)^{-1} = \frac{1}{\det A(z)} \tilde{A}(z)
 \]
where $\tilde{A}$ denotes the cofactor matrix of $A$.
 We assume that $1/\det A(z)$ is regular on the unit circle $C$ and 
has $2N$ isolated simple poles $\{z_1, z_2, \ldots, z_{2N} \}$ in
$|z|>1$ (see \eref{eq_ref7} for the reason why we choose the number of 
poles to be $2N$). We set
\[
J_{\infty, n} = \lim_{z \to \infty} \phi_n z^{-n} A(z)^{-1}
\]
and use the residue theorem. After some computation, 
we arrive at the discrete version of the Gel'fand-Levitan-Marchenko equation,
 \beq
 \bar{K}(n,m) +  \sum_{n'=n}^{\infty} K(n,n') F(n'+m)
         = J_{\infty, n} \de_{n,m}
 \hspace{3mm} (m \geq n).
 \label{GL1}
 \eeq
Here $F(n'+m)$ is defined by
 \beq
 F(n'+m) \equiv
 \frac{1}{2\pi \i} \oint_{C} B(z)A(z)^{-1} z^{-(n'+m)-1}
\d z + \sum_{j=1}^{2N} C_j z_j^{-(n'+m)-1}
 \label{F_form}
 \eeq
where $C_j$ is the residue matrix of $B(z) A(z)^{-1}$ at $z = z_j$. 

Similarly, we operate
 \[
 \frac{1}{2\pi \i}\oint_{C} \d z \; z^{m-1} \hspace{5mm} (m \ge n)
 \]
on both sides of \eref{prep2} substituting \eref{psi_form}. 
As has been mentioned previously,
 $\5{\phi}_n z^{n}$ and $\5{A}(z)$ are analytic inside the
unit circle $C$, $|z|<1$.
 We assume that $1/\det \5{A}(z)$ is regular on the unit 
circle $C$ and has $2\5{N}$ isolated simple
 poles $\{\5{z}_1, \5{z}_2, \ldots, \5{z}_{2\5{N}}\}$ in $|z|<1$. 
We set
\[
\bar{J}_{0, n} = \lim_{z \to 0} \bar{\phi}_n z^{n} \bar{A}(z)^{-1}
\]
and use the residue theorem. Finally, we obtain
 the counterpart of the discrete Gel'fand-Levitan-Marchenko equation,
 \beq
  K(n,m) - \sum_{n'=n}^{\infty} \bar{K}(n,n') \bar{F}(n'+m) = 
        -\bar{J}_{0,n} \de_{n,m}
 \hspace{3mm} (m \geq n).
 \label{GL2}
 \eeq
Here $\5{F}(n'+m)$ is defined by
 \beq
 \5{F}(n'+m) \equiv
\frac{1}{2\pi \i} \oint_{C} 
\bar{B}(z)\bar{A}(z)^{-1} z^{n'+m-1}\d z 
 - \sum_{k=1}^{2 \5{N}} \5{C}_k \5{z}_k^{n'+m-1}.
 \label{F_bar_form}
 \eeq
The matrix $\5{C}_k$ is the residue matrix of $\5{B}(z) \5{A}(z)^{-1}$
 at $z = \5{z}_k$. 
From \eref{K3} and \eref{K_bar3}, it is natural to set
\[
K(n,m) = \kappa(n,m) \prod_{i=n}^\infty (I-\ti{R}_i \ti{Q}_i)^{-1}
= \kappa (n,m) \prod_{i=n}^{\infty} \rho_i^{-1}
\hspace{5mm} (m \geq n)
\]
\[
\5{K}(n,m) = \5{\kappa}(n,m) \prod_{i=n}^\infty (I-\ti{Q}_i \ti{R}_i)^{-1}
= \5{\kappa} (n,m) \prod_{i=n}^{\infty} \rho_i^{-1}
\hspace{5mm} (m \geq n).
\]
Here $\kappa(n,m)$ and $\5{\kappa}(n,m)$ are column vectors
 whose elements are $l \times l$ square matrices,
 \[
 \kappa(n,m) =
 \left[
 \begin{array}{c}
  \kappa_1(n,m)  \\
  \kappa_2(n,m)  \\
 \end{array}
 \right]
\hspace{5mm}
 \bar{\kappa}(n,m) =
 \left[
 \begin{array}{c}
  \bar{\kappa}_1(n,m)  \\
  \bar{\kappa}_2(n,m)  \\
 \end{array}
 \right].
 \]
In particular, $\kappa(n,n)$ and $\bar{\kappa}(n,n)$ are given by
\[
\kappa (n,n) =
 \left[
 \begin{array}{c}
   O  \\
   I  \\
 \end{array}
 \right]
\hspace{5mm}
\5{\kappa} (n,n) =
 \left[
 \begin{array}{c}
   I  \\
   O  \\
 \end{array}
 \right].
\]
%
Due to \eref{K_eq} and \eref{K_bar_eq}, the potentials 
$\ti{Q}_n$ and $\ti{R}_n$ are given by
 \beq
 -\kappa_1(n,n+1) = \ti{Q}_n
 \label{ka1}
 \eeq
 \beq
 -\bar{\kappa}_2(n,n+1) = \ti{R}_n .
 \label{ka_bar1}
 \eeq
In terms of $\kappa$ and $\5{\kappa}$, the Gel'fand-Levitan-Marchenko 
equations \eref{GL1} and \eref{GL2} for $m>n$ are rewritten as
 \beq
\fl
 \bar{\kappa}(n,m) + 
 \left[
 \begin{array}{c}
   O  \\
   I  \\
 \end{array}
 \right] F(n+m)
 + \sum_{n'=n+1}^{\infty} \kappa(n,n') F(n'+m) = 
 \left[
 \begin{array}{c}
  O  \\
  O  \\
 \end{array}
 \right]
 \hspace{5mm} (m>n)
 \label{GL5}
 \eeq
 \beq
\fl
 \kappa(n,m) -
 \left[
 \begin{array}{c}
   I  \\
   O  \\
 \end{array}
 \right] \bar{F}(n+m)
 - \sum_{n'=n+1}^{\infty} \bar{\kappa}(n,n') \bar{F}(n'+m) = 
 \left[
 \begin{array}{c}
  O  \\
  O  \\
 \end{array}
 \right]
 \hspace{5mm} (m>n).
 \label{GL6}
 \eeq
 
It should be noted that the scattering problem
 \eref{scattering_problem2} gives the symmetry properties 
of the scattering data.
%
For instance, we have
\beq
\det A(z) = \det A(-z)
\hspace{5mm}
\det \5{A}(z) = \det \5{A}(-z)
\label{eq_ref7}
\eeq
which means that the eigenvalues $z_j$, $\5{z}_k$ should appear as 
`positive-negative' pairs. Further, we have 
\beq
\fl
B(z) A(z)^{-1} = -B(-z) A(-z)^{-1}
\hspace{5mm}
\5{B}(z) \5{A}(z)^{-1} = -\5{B}(-z) \5{A}(-z)^{-1}.
\label{eq_ref21}
\eeq
Therefore, we can simplify the forms of $F$ and $\5{F}$ as
\[
\eql{
F(n+m) = 
\left\{
\begin{array}{c}
 2 F_R (n+m) \\
  O \\
\end{array}
\right.
\begin{array}{l}
 m= n+2j-1  \\
 m=n+2j \\
\end{array}
\; \; \; j \geq 1
\\
F_R(n) = \frac{1}{2 \pi \i} \int_{C_R} 
B(z)A(z)^{-1} z^{-n-1} \d z+ \sum_{j=1}^N C_j z_j^{-n-1} 
}
\]
and
\[
\eql{
\5{F}(n+m) = 
\left\{
\begin{array}{c}
 2 \5{F}_R (n+m)\\
  O \\
\end{array}
\right.
\hspace{2mm}
\begin{array}{l}
 m= n+2j-1 \\
 m=n+2j \\
\end{array}
\; \; \; j \geq 1
\\
\5{F}_R(n) = \frac{1}{2 \pi \i} \int_{C_R} 
\5{B}(z)\5{A}(z)^{-1} z^{n-1} \d z - \sum_{k=1}^{\5{N}} \5{C}_k
 \5{z}_k^{n-1}.
}
\]
Here $C_R$ denotes the right-half portion of the 
unit circle contour $C$. 

The symmetry properties of $F$ and $\5{F}$ give rise to 
those of $\kappa$ and $\5{\kappa}$. From \eref{GL5} and 
\eref{GL6}, we obtain
\num
\bea
\kappa_1 (n,m) = 
\left\{
\begin{array}{c}
 \kappa_{1 R} (n,m) \\
  O \\
\end{array}
\begin{array}{l}
 m= n+2j-1  \\
 m=n+2j \\
\end{array}
\right.
\; \; \; j \geq 1
\\
\5{\kappa}_2 (n,m) = 
\left\{
\begin{array}{c}
  \5{\kappa}_{2 R} (n,m) \\
  O \\
\end{array}
\begin{array}{l}
 m= n+2j-1 \\
 m=n+2j  \\
\end{array}
\right.
\; \; \; j \geq 1.
\eea
\enum
Considering the above symmetry properties, we obtain the simplified 
Gel'fand-Levitan-Marchenko equations for $\kappa_{1 R}$ and 
$\5{\kappa}_{2 R}$,
\bea
\fl &\kappa_{1 R} (n,m) \hspace{5mm} (m>n, \, m-n = {\rm odd})
\nn \\
\fl =& 2 \5{F}_R(n+m) - 
        4\dsum{n'=n+2}{n'-n={\rm even}}^\infty 
        \dsum{n''=n+1}{n''-n={\rm odd}}^\infty 
         \kappa_{1 R} (n, n'') F_R(n''+n') \5{F}_R(n'+m)
\label{eq_ref23}
\eea
\bea
\fl & \5{\kappa}_{2 R} (n,m) \hspace{5mm} (m>n, \, m-n = {\rm odd})
\nn \\
\fl =& -2F_R(n+m) - 
        4\dsum{n'=n+2}{n'-n={\rm even}}^\infty 
        \dsum{n''=n+1}{n''-n={\rm odd}}^\infty 
         \5{\kappa}_{2 R} (n, n'') \5{F}_R(n''+n') F_R(n'+m).
\label{eq_ref24}
\eea

\subsection{Time dependence of the scattering data}
 \label{Time-dep}

 Under the rapidly decreasing boundary conditions \eref{boundary},
 the asymptotic form of the 
Lax matrix $\ti{M}_n$ for the sd-CNLS 
equations \eref{CNLS4} after the gauge transformation \eref{gauge} 
is given by
 \[
 \ti{M}_n = g_n M_n g_n^{-1} + g_{n,\, t}g_n^{-1} \rightarrow 
 \left[
 \begin{array}{cc}
  z^2 I  &  O \\
   O  &  \frac{1}{z^2} I \\
 \end{array}
 \right]
 \hspace{5mm} {\rm as} \hspace{4mm} n \rightarrow \pm \infty .
 \]

We define time-dependent Jost functions by
 \[
\fl
 \phi_n^{(t)} \equiv \phi_n \e^{z^2 t} \sim
 \left[
 \begin{array}{c}
  I \\
  O \\
 \end{array}
 \right]
 z^n \e^{z^2 t}
\hspace{5mm}
 \bar{\phi}_n^{(t)} \equiv \bar{\phi}_n \e^{\frac{1}{z^2} t} \sim
 \left[
 \begin{array}{c}
  O \\
  -I \\
 \end{array}
 \right]
 z^{-n} \e^{\frac{1}{z^2} t}
 \hspace{7mm}
 {\rm as}~~~ n \rightarrow -\infty .
 \]
From the relations
 \[
 \phi_{n, t}^{(t)} = \ti{M}_n \phi_n^{(t)} \hspace{5mm}
 \bar{\phi}_{n, t}^{(t)} = \ti{M}_n \bar{\phi}_n^{(t)}, 
 \]
we obtain
 \beq
 \phi_{n, t} = (\ti{M}_n - z^2 I) \phi_n
 \hspace{5mm} \bar{\phi}_{n, t} = \bigl(\ti{M}_n -
\frac{1}{z^2} I\bigr) \bar{\phi}_n.
 \label{time1}
 \eeq

We put the definitions of the time-dependent scattering data,
\num
 \bea
 \phi_n(z) = \bar{\psi}_n(z)A(z, t) + \psi_n(z)B(z, t)
\nn
\\
 \bar{\phi}_n(z) = \bar{\psi}_n(z) \bar{B}(z, t) - \psi_n(z)
 \bar{A}(z, t)
\nn
 \eea
\label{time_total}
\enum
into \eref{time1}. Then taking the limit 
 $n \rightarrow +\infty$, we 
obtain the time dependences of $A$, $BA^{-1}$, $C_j$ and $\bar{A}$, 
$\5{B} \5{A}^{-1}$, $\bar{C}_k$. They are given by, respectively,
\[
A(z, t) = A(z, 0)
\]
 \beq
\fl
 B(z,t) A(z, t)^{-1} = B(z, 0)A(z, 0)^{-1}
\e^{-\bigl(z^2-\frac{1}{z^2}\bigr) t}
\hspace{5mm}
 C_j(t) = C_j(0)\e^{-\bigl(z_j^2-\frac{1}{z_j^2}\bigr) t}
 \label{C_time}
 \eeq
 and
\[
\bar{A}(z, t) = \bar{A}(z, 0)
\]
\beq
\fl
 \5{B}(z,t) \5{A}(z, t)^{-1} 
 = \5{B}(z, 0) \5{A}(z, 0)^{-1} \e^{\bigl(z^2-\frac{1}{z^2}\bigr) t}
\hspace{5mm}
 \bar{C}_k(t) = \bar{C}_k(0)
\e^{\bigl(\bar{z}_k^2-\frac{1}{\5{z}_k^2}\bigr) t}.
 \label{C_bar_time}
 \eeq
The above results give explicitly time-dependent
 forms of $F_R(n, t)$ and $\5{F}_R(n, t)$ for odd $n$.
 %
 %

 \subsection{Initial-value problem}
 \label{Initial value problem}
Thanks to the constraints $\ti{R}_n=-\ti{Q}_n^{\, \dagger}$ and
 $\ti{Q}_n \ti{R}_n = \ti{R}_n \ti{Q}_n = -\sigma_n I$, 
we have some additional relations besides  
\eref{eq_ref7} and \eref{eq_ref21}. The first additional relation is
 \beq
 \det \5{A}(z) = \bigl\{\det A \bigl(\frac{1}{z^{\ast}}\bigr)\bigr\}^{\ast}
 \label{det_A2}
 \eeq
which is proved in \cite{Tsuchida2}. This relation restricts
 the numbers and the positions of the poles of $1/\det A(z)$ and
 $1/\det \5{A}(z)$, i.e.
 \beq
 \5{N} = N \hspace{5mm} \5{z}_k= \frac{1}{z_k^{\ast}}.
 \label{eq_ref22}
 \eeq

Due to \eref{scat_rel3}, we have the second additional relations,
\bseq
 \bea
 \5{B}(z)\5{A}(z)^{-1} = \{ B(z) A(z)^{-1}\}^{\dagger}
 \hspace{2mm} ( {\rm on} \; |z|=1)
\\
 \5{C}_k = -\frac{1}{z_k^{\ast\,2}} C_k^{\, \dagger}.
 \eea
 \label{scat_rel4}
\eseq
The relations \eref{eq_ref22} and 
\eref{scat_rel4} give a relation between
 $\5{F}_R(n, t)$ and $F_R(n, t)$,
 \beq
 \5{F}_R(n,t) = F_R(n,t)^{\dagger}.
 \label{F_bar_F}
 \eeq
In order to make the ISM applicable to the 
sd-CNLS equations,
 we have to take account of the internal symmetries
 of the potentials 
$\ti{Q}_n$ and $\ti{R}_n$. 
According to \eref{QR_def1}, \eref{QR_def3} and \eref{QR_trans}, 
the potentials $\ti{Q}_n$ and $\ti{R}_n$ are defined recursively by
\[
\ti{Q}^{(1)}_n = \i^n \e^{2\i t} q^{(1)}_n 
\hspace{10mm} \ti{R}^{(1)}_n = (-\i)^n \e^{-2\i t} r^{(1)}_n
\]
\[
 \ti{Q}^{(m+1)}_n
 = 
 \left[
 \begin{array}{cc}
  \ti{Q}^{(m)}_n &  \i^n \e^{2\i t} q^{(m+1)}_n I_{2^{m-1}} \\
  (-\i)^n \e^{-2\i t} r^{(m+1)}_n I_{2^{m-1}} &  -\ti{R}^{(m)}_n \\
 \end{array}
 \right]
\]
 \[
 \ti{R}^{(m+1)}_n
 = 
 \left[
 \begin{array}{cc}
  \ti{R}^{(m)}_n &  \i^n \e^{2\i t} q^{(m+1)}_n I_{2^{m-1}} \\
  (-\i)^n \e^{-2\i t} r^{(m+1)}_n I_{2^{m-1}} &  -\ti{Q}^{(m)}_n \\
 \end{array}
 \right].
 \]

If we set 
\beq
\eql{
\i^n \e^{2\i t} q_n^{(j)} = v_n^{(2j-2)} + \i v_n^{(2j-1)}
\\
(-\i)^n \e^{-2\i t} r_n^{(j)} = -v_n^{(2j-2)} + \i v_n^{(2j-1)}
}
\hspace{5mm} j=1,2, \ldots, m
\eeq
$\ti{Q}_n^{(m)}$ and $\ti{R}_n^{(m)}$ for $m\ge 2$ are written as
 \beq
 \ti{Q}_n^{(m)}= v_n^{(0)} \eins + \sum_{k=1}^{2m-1} v_n^{(k)} e_k
 \hspace{5mm}
 \ti{R}_n^{(m)}= -v_n^{(0)} \eins + \sum_{k=1}^{2m-1} v_n^{(k)} e_k .
 \label{Q_R_const}
 \eeq
%
Here $\eins$ is 
the $2^{m-1}\times 2^{m-1}$ unit matrix, which has been denoted by 
$I_{2^{m-1}}$. Because of $r_n^{(j)} = -q_n^{(j) \ast}$, 
$v_n^{(0)}$ and $v_n^{(k)}$ should be real. 
Substitution of \eref{Q_R_const} into the relations \eref{eq_ref91} 
yields the following important 
relations for $2^{m-1}\times 2^{m-1}$ matrices $\{e_i\}$:
 \bea
 \{ e_i, e_j \}_+ = -2 \de_{i,j} \eins
 \label{ei_ej}
 \\
 e_k^{\, \dagger} = - e_k . 
 \label{ek_dag}
 \eea
Here $\{\cdot, \cdot \}_+$ denotes the anti-commutator. 
It is stressed that expressions \eref{Q_R_const} with 
conditions \eref{ei_ej} and \eref{ek_dag} are only a sufficient 
(i.e. not necessary) condition of \eref{eq_ref91}. 

Considering the above symmetries of the potentials 
$\ti{Q}_n^{(m)}$ and $\ti{R}_n^{(m)}$ for $m \ge 2$, we 
can show the following properties of the scattering data. \\
{\bf Proposition 3.$\,$4.$\,$1} \\
(1) The determinant of $A(z)$ and the determinant of $A(z^\ast)$ are 
related by
\beq
\det {A}(z) = \{\det A (z^{\ast})\}^{\ast}.
\label{A_res}
\eeq
Thus the poles of $1/\det A(z)$ in $|z| >1$ 
appear as pairs situated symmetric 
with respect to the real axis. Therefore, we replace $2N$ in section 
3.$\,$2 with $4N$ and choose the values of $2N$ poles in $|z|>1$, ${\rm Re}\, z
\geq 0$ as
\bea
\eql{
z_{2j-1} = \xi_j + \i \eta_j = a_j \e^{\i \theta_j}
\\
z_{2j} = z_{2j-1}^\ast = \xi_j - \i \eta_j = a_j \e^{-\i \theta_j}
}
 \hspace{5mm} & j=1, 2, \ldots, N
\label{z_z}
\eea
where $a_j > 1$, $0 < \theta_j \leq \pi/2$ for $\theta_j \neq 0$.
The conditions \eref{z_z} should be 
interpreted as follows. If $\theta_j = 0$, the corresponding pole
 does not need its counterpart. The values of the remaining 
$2N$ poles in $|z|>1$, ${\rm Re}\, z \leq 0$ are given by
\[
z_{2N+k} = -z_k \hspace{5mm} k=1,2,\ldots, 2N.
\]
(2) The reflection coefficient 
$B(z) A(z)^{-1}$ is expressed as
\beq
B(z) A(z)^{-1} = r^{(0)} \eins + \sum_{k=1}^{2m-1}r^{(k)} e_k.
\label{}
\eeq
Here $r^{(0)}$ and $r^{(k)}$ are complex functions of $z$ and $t$ 
which satisfy
\beq
r^{(0)}(z^\ast) = 
r^{(0)}(z)^{\ast} \hspace{5mm}
r^{(k)}(z^\ast) = 
r^{(k)}(z)^{\ast}.
\label{r_r}
\eeq
%
(3) The residue matrices $\{ C_1, C_2, \ldots, C_{2N}\}$ are expressed as
\beq
\eql{
C_{2j-1} = c_j^{(0)} \eins + \sum_{k=1}^{2m-1} c_j^{(k)} e_k
\\
C_{2j} = c_j^{(0)\, \ast} \eins 
+ \sum_{k=1}^{2m-1} c_j^{(k)\, \ast} e_k
}
\hspace{5mm} j=1, 2, \ldots, N
\label{C_C}
\eeq
with complex functions of time $t$, $c_j^{(0)}$ and $c_j^{(k)}$. 
%
\\
The statements (1)--(3) are proved 
essentially in the same way as in \cite{Tsuchida2} and 
therefore their proofs are omitted. 

Taking account of the above conditions, we obtain explicit 
expressions of $F_R(n,t)$ and $\5{F}_R(n,t)$ for odd $n$, 
\bea
\fl  F_R (n, t) &=& \frac{1}{2\pi \i} \int_{C_R}
        B(z) A(z)^{-1} z^{-n-1} \d z 
         + \sum_{j=1}^{2N} C_j z_j^{-n-1}
\nn 
\\
\fl &=&
\frac{1}{2\pi \i} \int_{C_{UR}}
        \Bigl\{ (r^{(0)} z^{-n-1}+r^{(0)\, \ast} z^{n-1}) \eins
         + \sum_{k=1}^{2m-1}
        ( r^{(k)} z^{-n-1} + r^{(k)\,\ast} z^{n-1}) e_k \Bigr\}
        \, \d z 
\nn \\
\fl &&   + \sum_{j=1}^{N} \Bigl\{ (c_j^{(0)} z_j^{-n-1}
        +c_j^{(0)\, \ast} z_j^{\ast\, -n-1})\eins
         + \sum_{k=1}^{2m-1} (c_j^{(k)}z_j^{-n-1}
         +c_j^{(k)\, \ast}z_j^{\ast\, -n-1}) e_k \Bigr\} 
\label{F_form3}
\eea
\bea
\fl  \5{F}_R(n, t) &=& F_R(n, t)^{\dagger} 
\nn \\
\fl &=&
\frac{1}{2\pi \i} \int_{C_{UR}}
        \Bigl\{ (r^{(0)} z^{-n-1}+r^{(0)\, \ast} z^{n-1}) \eins
         - \sum_{k=1}^{2m-1}
        ( r^{(k)} z^{-n-1} + r^{(k)\, \ast} z^{n-1}) e_k \Bigr\}
        \, \d z 
\nn \\
\fl &&   + \sum_{j=1}^{N} \Bigl\{ (c_j^{(0)} z_j^{-n-1}
        +c_j^{(0)\, \ast} z_j^{\ast\, -n-1})\eins
         - \sum_{k=1}^{2m-1} (c_j^{(k)}z_j^{-n-1}
         +c_j^{(k)\, \ast}z_j^{\ast\, -n-1}) e_k \Bigr\} 
\label{F_bar_form3}
\eea
where $C_{UR}$ denotes a contour along the quadrant (upper-right portion)
 of the unit circle $C$. 
We see that the coefficients of $\eins$ and $\{ e_k \}$ in 
\eref{F_form3} and \eref{F_bar_form3} are real. 
%
Thus, a pair of $\5{F}_R$ and $-F_R$ is expressed in the same form 
as \eref{Q_R_const}, as is expected from the viewpoint of successive 
approximations for the Gel'fand-Levitan-Marchenko equations. 

Because $B(z)A(z)^{-1}$ and $C_j$ depend on $t$ as \eref{C_time}, 
the time dependences of $r^{(0)}$, $r^{(k)}$
 and $c_j^{(0)}$, $c_j^{(k)}$ are given by
\beq
 r^{(0)}(z,t) = r^{(0)}(z,0) \e^{-\bigl(z^2-\frac{1}{z^2}\bigr) t}\hspace{5mm}
 r^{(k)}(z,t) = r^{(k)}(z,0) \e^{-\bigl(z^2-\frac{1}{z^2}\bigr) t}
 \label{r_time}
 \eeq
 \beq
 c_j^{(0)}(t) = c_j^{(0)}(0) \e^{-\bigl(z_j^2-\frac{1}{z_j^2}\bigr) t}\hspace{5mm}
 c_j^{(k)}(t) = c_j^{(k)}(0) \e^{-\bigl(z_j^2-\frac{1}{z_j^2}\bigr) t}.
 \label{c_time}
 \eeq
%
Combining \eref{eq_ref23} and \eref{eq_ref24} with \eref{F_bar_F},
 we arrive at
\bea
\fl & \kappa_{1 R} (n,m ;t) 
\nn \\
\fl =& 2F_R(n+m ,t)^{\dagger} 
- 4 \sum_{l_1=1}^\infty \sum_{l_2=1}^\infty 
         \kappa_{1 R} (n, n+2l_2-1;t) F_R(2n+2l_2+2l_1-1,t) 
        F_R(n+2l_1 +m,t)^\dagger
\nn \\
\fl &
\label{GL3}
\eea
 \bea
\fl
& \5{\kappa}_{2 R} (n,m;t)\hspace{5mm} 
\nn \\
\fl
=& -2F_R(n+m,t) 
-4 \sum_{l_1=1}^\infty \sum_{l_2=1}^\infty 
         \5{\kappa}_{2 R} (n, n+2l_2-1;t) F_R(2n+2l_2+2l_1-1,t)^\dagger
 F_R(n+2l_1 +m,t)
\nn \\
\fl
& \label{GL4}
\eea
for $m>n$, $m-n={\rm odd}$, where $F_R(n,t)$ is given by \eref{F_form3}. 
 %

Now the initial-value problem of the sd-CNLS equations
 \eref{CNLS4} can be solved in the following steps. \\
 (1) For given potentials at $t=0$, $\ti{Q}_n(0)$ and $\ti{R}_n(0)$ which are
expressed as \eref{Q_R_const}, we solve the 
 scattering problem \eref{scattering_problem2}, and obtain 
 the scattering data $ \{ B(z)A(z)^{-1}, z_j, C_j \}$ or, more concretely 
$\{ r^{(0)}, r^{(k)}, z_j, c^{(0)}_j, c^{(k)}_j\}$.
\\
 (2) The time dependence of the scattering data is given by 
 \eref{C_time} (or \eref{r_time} and 
\eref{c_time}).
 \\
 (3) We substitute the time dependence of the scattering data into
 the Gel'fand-Levitan-Marchenko equations \eref{GL3} and \eref{GL4}. 
 Solving the equations, we reconstruct the time-dependent potentials,
 \[
 \ti{Q}_n(t) = -\kappa_{1 R}(n,n+1;t)
 \]
 \[
 \ti{R}_n(t) = -\5{\kappa}_{2 R}(n,n+1;t).
 \]
This solution proves directly the complete integrability of the
sd-CNLS equations \eref{CNLS4}. 

\subsection{Soliton solutions}
\label{}
To construct soliton solutions of the sd-CNLS equations, 
we assume the reflection-free condition, i.e. $B(z)=\5{B}(z)= O$ on $|z|=1$. 
Then, $F_R(n,t)$ and $\5{F}_R(n,t)$ for odd $n$ are given by
 \beq
 F_R(n,t) = \sum_{j=1}^{2N} C_j(t) z_j^{-n-1}
  \hspace{7.0mm} C_j(t)= C_j(0) \e^{-\bigl(z_j^2-\frac{1}{z_j^2}\bigr) t}
 \label{F_ref_less}
 \eeq
 \beq
 \5{F}_R (n,t) = -\sum_{k=1}^{2N} \5{C}_k(t) \5{z}_k^{n-1}
  \hspace{5mm} \5{C}_k(t)= \5{C}_k(0) \e^{\bigl(\5{z}_k^2-\frac{1}{\5{z}_k^2}\bigr) t}.
 \label{F_bar_less}
 \eeq
To solve \eref{eq_ref23} (or equation \eref{GL3})
  with \eref{F_ref_less} and \eref{F_bar_less}, we set
  \beq
  \kappa_{1 R} (n, m; t) = \sum_{k=1}^{2N} P_k \5{C}_k(t) 
   \5{z}_k ^{n+m-1} \hspace{5mm} (m-n={\rm odd}).
  \label{K_1_form}
  \eeq
 Substituting \eref{K_1_form} into \eref{eq_ref23} or 
\eref{GL3}, we have
 \beq
 P_k  -4 \sum_{l=1}^{2N} \sum_{j=1}^{2N} 
 \Bigl( \frac{\5{z}_l}{z_j} \Bigr)^{2n}
\frac{\5{z}_k^2} {(z_j^2 -\5{z}_k^2)(z_j^2-\5{z}_l^2)} P_l 
 \5{C}_l(t) C_j (t) = - 2I.
 \label{P_eq}
 \eeq
In terms of a matrix $S$ whose elements are defined by
 \bea
\fl  S_{lk} &\equiv
\de_{l,k} I -4 \sum_{j=1}^{2N} 
\Bigl( \frac{\5{z}_l}{z_j} \Bigr)^{2n}
\frac{\5{z}_k^2}{(z_j^2 -\5{z}_k^{2})(z_j^2 - \5{z}_l^{2})}
        \5{C}_l(t) C_j(t)
\nn \\
\fl &=
\de_{l,k} I +4 \sum_{j=1}^{2N} 
 \frac{1}
 {z_j^{2n} z_l^{\ast\, 2n} (z_j^2 z_k^{\ast\, 2}-1)(z_j^2 z_l^{\ast\, 2}-1)}
C_l(t)^\dagger C_j(t)
 \hspace{5mm} 1 \le l,k \le 2N 
\nn
 \eea
equation \eref{P_eq} is expressed by 
 \beq
 (\; P_1 \; P_2 \; \cdots \; P_{2N} \; )
 \left(
 \begin{array}{ccc}
  S_{11} & \cdots & S_{1\,2N}\\
  \vdots & \ddots & \vdots\\
  S_{2N\, 1} & \cdots & S_{2N\, 2N} \\
 \end{array}
 \right)
 =
 - 2(\, \underbrace{\, I \; I \; \cdots \; I \,}_{2N} \, ).
 \label{mat_eq1}
 \eeq
Similarly, we solve \eref{eq_ref24} (or equation \eref{GL4}) with 
 \eref{F_ref_less} and \eref{F_bar_less}. Substitution of
 \beq
 \5{\kappa}_{2 R} (n, m; t) = \sum_{j=1}^{2N} \5{P}_j C_j(t) 
  z_j^{-(n+m)-1} \hspace{5mm} (m-n={\rm odd})
 \label{Kbar_2_form}
 \eeq
into \eref{eq_ref24} or \eref{GL4} gives
 \beq
 \5{P}_j -4 \sum_{l=1}^{2N} \sum_{k=1}^{2N} 
 \frac{\5{z}_k^{2n+2}}{z_l^{2n}}
\frac{1}{(z_l^2-\5{z}_k^2)(z_j^2 -\5{z}_k^2)} \5{P}_l 
 C_l(t) \5{C}_k (t) = -2I.
 \label{P_bar_eq}
 \eeq
Using a matrix $\5{S}$,
 \bea
\fl \5{S}_{lk} &\equiv
\de_{l,k}I -4 \sum_{j=1}^{2N} 
 \frac{\5{z}_j^{2n+2}}
 {z_l^{2n}}\frac{1}{(z_l^2 - \5{z}_j^2)(z_k^2 - \5{z}_j^{2})}
        C_l(t) \bar{C}_j(t)
\nn \\
\fl &=
\de_{l,k}I +4 \sum_{j=1}^{2N} 
 \frac{1}
 {z_l^{2n} z_j^{\ast\, 2n} (z_l^2 z_j^{\ast\, 2}-1)(z_k^2 z_j^{\ast\, 2}-1)}
C_l(t)C_j(t)^\dagger
 \hspace{5mm} 1 \le l,k \le 2N
\nn
 \eea
we rewrite \eref{P_bar_eq} as
 \beq
 (\; \5{P}_1 \; \5{P}_2 \; \cdots \; \5{P}_{2N} \; )
 \left(
 \begin{array}{ccc}
  \5{S}_{11} & \cdots & \5{S}_{1\,2N}\\
  \vdots & \ddots & \vdots\\
  \5{S}_{2N\, 1} & \cdots & \5{S}_{2N\, 2N} \\
 \end{array}
 \right)
 =
 -2(\, \underbrace{\, I \; I \; \cdots \; I \,}_{2N} \, ).
 \label{mat_eq2}
 \eeq
Equations \eref{mat_eq1} and \eref{mat_eq2} are readily solved. 
Thus the $N$-soliton solution of the sd-CNLS equations
 \eref{CNLS4} is given by 
\begin{subequations}
 \bea
 \ti{Q}^{(m)}_n(t) 
 &=& \i^n (F_1)^n \e^{-\i H_1 t} Q_n^{(m)} \e^{\i H_2 t} (F_2)^n
\nn \\
&=& - \kappa_{1 R}(n,n+1;t)
 \nn \\
 &=& - 2\, (\, \underbrace{\, I \; I \; \cdots \; I \,}_{2N} \, )
 \; S^{-1} \;
 \left(
 \begin{array}{c}
\displaystyle
  C_1 (t)^{\dagger} \frac{1}{z_1^{\ast \, 2n+2}} \\
\displaystyle
  C_2 (t)^{\dagger} \frac{1}{z_2^{\ast \, 2n+2}} \\
\displaystyle
    \vdots  \\
\displaystyle
  C_{2N} (t)^{\dagger} \frac{1}{z_{2N}^{\ast \, 2n+2}} \\
 \end{array}
 \right)
 \label{N-soliton1}
\\
 \ti{R}^{(m)}_n(t) &=& 
(-\i)^n (F_2)^n \e^{-\i H_2 t} R_n^{(m)} \e^{\i H_1 t} (F_1)^n
\nn \\
&=& - \5{\kappa}_{2 R}(n,n+1;t)
 \nn \\
 &=& 2 \, (\, \underbrace{\, I \; I \; \cdots \; I \,}_{2N} \, )
 \; \bar{S}^{-1} \;
 \left(
 \begin{array}{c}
\displaystyle
  C_1 (t) \frac{1}{z_1^{2n+2}} \\
\displaystyle
  C_2 (t) \frac{1}{z_2^{2n+2}} \\
\displaystyle
    \vdots  \\
\displaystyle
  C_{2N} (t) \frac{1}{z_{2N}^{2n+2}} \\
 \end{array}
 \right).
 \label{N-soliton2}
 \eea
\label{eq_ref41}
\end{subequations}
%
Strictly speaking, equation \eref{eq_ref41} includes breathers besides
solitons. In order to extract pure soliton solutions, we assume 
that each soliton seen in $\sum_j |q_n^{(j)}(t)|^2$ has a
time-independent shape. By calculating an asymptotic behavior of the
tails of solitons at $n \, \to +\infty$, we obtain the corresponding
necessary conditions
\beq
\fl
C_{2j-1} \bar{C}_{2j} = \bar{C}_{2j} C_{2j-1}
 = C_{2j} \bar{C}_{2j-1} = \bar{C}_{2j-1} C_{2j} = O
\hspace{10mm} j=1, 2, \ldots, N
\label{C_bar_C}
\eeq
on the residue matrices. Equation \eref{C_bar_C} is translated explicitly into 
\beq
\sum_{i=0}^{2m-1} (c^{(i)}_j)^2 = 0
\hspace{10mm} j=1, 2, \ldots, N .
\label{c^2}
\eeq
As an example, we write down a pure one-soliton solution. Choose 
$N=1$ and set
\bea
\bar{z}_1 &=& \frac{1}{z_1^\ast} = \e^{-W+\i \theta}
\hspace{5mm} W>0
\nn \\
\bar{C}_1 &=& -\frac{1}{z_1^{\ast \, 2}}C_1^{\, \dagger}
= -\frac{1}{z_1^{\ast \, 2}}
        \Bigl( c_1^{(0)\, \ast} \eins - \sum_{k=1}^{2m-1} 
        c_1^{(k)\, \ast} e_k \Bigr)
\nn \\
& \equiv & \bar{c}_1^{(0)} \eins + \sum_{k=1}^{2m-1} \bar{c}_1^{(k)} e_k 
\nn \\
\bar{C}_2 &=& \bar{c}_1^{(0)\, \ast} \eins
 + \sum_{k=1}^{2m-1} \bar{c}_1^{(k)\, \ast} e_k 
\nn \\
\e^{\phi_0} &=& 
\frac{\sinh 2W}{\sqrt{\displaystyle 2\sum_{j=0}^{2m-1}
|\bar{c}_1^{(j)}(0)|^2}} \, .
\label{phi_0}
\eea
Then, from \eref{eq_ref41} we obtain
\bseq
\begin{eqnarray}
\fl
\ti{Q}_n^{(m)} (t) &=& 
  \sech \bigl\{ 2 n W + 2(\sinh 2W \cos 2\theta) t + \phi_0 \bigr\} 
   \frac{\sinh 2W}{\sqrt{\displaystyle 2\sum_{j=0}^{2m-1}|\bar{c}_1^{(j)}(0)|^2}}
\nn \\  
\fl
&&
  \cdot \bigl\{ \bar{C}_1 (0)   \e^{2\i \{n \theta + (\cosh 2W \sin 2\theta) t \}}
   + \bar{C}_2 (0) \e^{-2\i \{n \theta + (\cosh 2W \sin 2\theta) t \}} 
  \bigr\}
\\
\fl 
\ti{R}^{(m)}_n (t) &=& -\ti{Q}_n^{(m)}(t)^\dagger .
\end{eqnarray}
\label{1-soli}
\eseq
It is straightforward to show that \eref{1-soli} can be expressed as 
\eref{Q_R_const} with real coefficients $v_n^{(i)}(t)$ of $\eins$ and 
$\{ e_k \}$. Thus we have checked in terms of the inverse problem that
the conditions \eref{Q_R_const} or consequently \eref{eq_ref91} are 
satisfied under proposition 3.$\,$4.$\,$1 and the conditions 
\eref{C_bar_C}, in the case of the one-soliton solution.

By introducing a new set of constants by
\[
\eql{
\alpha_i \equiv \bar{c}_1^{(2i-2)}(0) + \i \bar{c}_1^{(2i-1)}(0)
\\
\beta_i  \equiv \bar{c}_1^{(2i-2)}(0) - \i \bar{c}_1^{(2i-1)}(0)
}
\hspace{10mm} i=1,2, \ldots, m 
\]
we obtain a simplified expression of the pure one-soliton solution 
of the sd-CNLS equations \eref{CNLS4}, namely,
\bea
\fl 
q_n^{(i)}(t) 
&=&
\sech \bigl\{ 2 n W + 2(\sinh 2W \cos 2\theta) t + \phi_0 \bigr\} 
   \frac{\sinh 2W}{\sqrt{\displaystyle \sum_{j=1}^{m}
(|\alpha_j|^2 +|\beta_j|^2)}}
  \nn \\  
\fl 
&&
  \cdot \bigl[ \alpha_i \e^{2\i \{n (\theta - \frac{\pi}{4}) 
   + (\cosh 2W \sin 2\theta -1) t \}}
   + \beta_i^\ast \e^{-2\i \{n (\theta+\frac{\pi}{4}) 
   + (\cosh 2W \sin 2\theta +1) t \}} 
  \bigr] 
\nn \\
\fl 
&& \hspace{60mm} i=1, 2, \ldots, m .
\label{1-solit}
\eea
Here the condition \eref{c^2} for $N=1$ is
 cast into the orthogonality condition,
\[
\sum_{i=1}^{m} \alpha_i \beta_i =0.
\]
The soliton solution \eref{1-solit} exhibits a novel property as 
a solution of NLS-type equations. Because there are two carrier waves 
in one envelope soliton, the shape of soliton
observed in $|q_n^{(i)} (t)|^2$ periodically oscillates in
time. It is observed for \eref{1-solit} that the summation of
$|q_n^{(i)} (t)|^2$ with respect to components, $i(=1,2, \ldots, m)$ 
\[
\sum_{i=1}^{m} |q_n^{(i)}(t)|^2 = 
\sinh^2 2W \sech^2 \bigl\{ 2 n W + 2(\sinh 2W \cos 2\theta) t + \phi_0
\bigr\} 
\]
has a time-independent shape, as is expected.
This fact suggests that the conditions \eref{C_bar_C} are necessary
and sufficient to give pure soliton solutions even in the $N$-soliton case. 

The continuum limit which reduces \eref{1-solit} into the one-soliton
solution of the continuous CNLS equations can be seen as follows. 
We denote by $h$ the (dimensionless) lattice spacing of the sd-CNLS
model. We rescale $t$ by 
\[
t \to \frac{1}{h^2} t
\]
and set
\bea
q_n^{(i)} (t) = h u_i (x,t) \hspace{5mm} x = n h
\nn \\
W = h \eta  \hspace{6mm} \theta -\frac{\pi}{4} = -h \xi
\nn \\
\alpha_i = h \gamma_i \hspace{6mm} \beta_i =0 .
\nn
\eea
If we take the continuum limit $h \to 0$, the one-soliton
solution \eref{1-solit} with \eref{phi_0} is transformed into 
\[
\fl
u_i (x, t) = 2 \eta \sech \{ 2\eta x + 8 \eta \xi t + \phi_0 \}
\frac{\gamma_i}{\sqrt{\displaystyle \sum_{j=1}^m |\gamma_j|^2}} 
\e^{-2\i \xi x - 4\i (\xi^2-\eta^2)t}
\hspace{5mm} i=1,2, \ldots, m
\]
with
\[
\e^{\phi_0} = \frac{2\eta}{\sqrt{\displaystyle \sum_{j=1}^m |\gamma_j|^2}} .
\]
This is indeed the one-soliton solution of the continuous CNLS equations,
\beq
\i \frac{\6 u_i}{\6 t} + \frac{\6^2 u_i}{\6 x^2} 
+ 2 \sum_{j=1}^m |u_j|^2 \cdot u_i = 0 
\hspace{5mm} i=1,2, \ldots, m.
\label{cCNLS}
\eeq
It is noteworthy that either $\alpha_i=0 \; (i=1,2,\ldots,m)$ or 
$\beta_i=0 \; (i=1,2,\ldots,m)$ is necessary for us to take the
continuum limit. This reflects the fact that the pure $N$-soliton
solution of the sd-CNLS equations \eref{CNLS4} includes more
arbitrary constants than the $N$-soliton solution of the continuous
CNLS equations \eref{cCNLS}. It has its origin in the oscillation of 
solitons seen in each component $|q_n^{(i)}(t)|^2$. In this sense, 
the structure of the pure soliton solution may be more similar to that
of the cmKdV equations \eref{cmKdV1} rather than to that of the CNLS
equations \eref{cCNLS}. 

Ohta \cite{Ohta} obtained an $N$-soliton solution for the sd-CNLS
equations in the Pfaffian representations. 
The $N$-soliton solution
\eref{eq_ref41} with the constraints \eref{C_bar_C} contains more
 parameters than Ohta's $N$-soliton solution. 
Therefore, it is reasonable to
 conjecture that our solution reduces to Ohta's 
by a particular choice of those parameters. 

We wish to show that \eref{N-soliton1} and \eref{N-soliton2} 
are expressed as 
\bseq
 \bea
 \ti{Q}^{(m)}_n(t)= v^{(0)}_n(t) \eins + \sum_{k=1}^{2m-1} v^{(k)}_n(t) e_k
 \label{Q_1_ei}
\\
 \ti{R}^{(m)}_n(t)= -v^{(0)}_n(t) \eins + \sum_{k=1}^{2m-1} v^{(k)}_n(t) e_k
 \label{R_1_ei}
 \eea
\label{QR_ei}
\eseq
without using the products of $\{ e_i \}$ such as 
$e_i e_j$, $e_i e_j e_k$. Further, we expect 
$v^{(0)}_n$, $v^{(k)}_n$ to be 
real in \eref{Q_1_ei} and \eref{R_1_ei}. As mentioned previously, 
`\eref{QR_ei} $\rightarrow$ \eref{eq_ref91}' holds. Thus, 
equation \eref{eq_ref91} is automatically satisfied if we can show
\eref{QR_ei}. So far we have 
proved  either \eref{Q_1_ei} or \eref{R_1_ei} only for $m =
2$. However, the result for the one-soliton solution implies that both of 
\eref{QR_ei} are simultaneously satisfied under proposition 3.$\,$4.$\,$1.

%

\section{Discussions}
\label{Discussions}

We have investigated the semi-discrete 
coupled nonlinear Schr\"{o}dinger (sd-CNLS) equations. The analysis
 from the ISM point of view is given 
for the first time in this paper. A previous paper 
\cite{Tsuchida2} dealt with a new extension of the semi-discrete
version of the ISM proposed by Ablowitz and Ladik to solve the 
semi-discrete coupled modified KdV (sd-cmKdV) equations \eref{cmV1}. 
In this paper, we have developed
 the extension with the help of the transformation \eref{trans} to
solve the model considered with arbitrarily multiple components. 

We should comment on the Lax formulations and the ISM 
for both the sd-cmKdV equations and the sd-CNLS equations,
which may spotlight mysterious structure of discrete soliton
equations. 

First, there is an essential difference between the form of the 
$L_n$-matrix for the sd-CNLS equations and that for the sd-cmKdV
equations in the case of $m \geq 2$ ($M \geq 4$). It seems that the
scattering problem for the sd-CNLS equations associated with
the $L_n$-matrix, e.g. \eref{L_n_example}, 
does not agree with the scattering
problem for the CNLS equations \cite{Manakov,Fordy} in the continuum
limit ($z=\e^{-\i \z h +\i \varphi}, \; h \to 0$). 
The situation is not observed
for the continuum limit of the sd-cmKdV equations. This difference
 can be understood by considering the continuum limit of the semi-discrete
coupled Hirota equations. 
A detailed explanation will be reported in a subsequent paper.

Secondly, it is obvious that \eref{mNLS1} does not generally 
allow us to assume 
the reduction $R_n=\pm Q_n^{\, \dagger}$, because of the order of the
products. To consider the
reduction $R_n=\pm Q_n^{\, \dagger}$, we should impose the additional 
restriction $Q_n R_n = R_n Q_n = {\rm scalar}$ on $Q_n$ and $R_n$. Thus, both 
\eref{eq_ref2} and \eref{R_Q} play a crucial role in our theory, which
is peculiar to the discrete theory with multiple components. 

Thirdly, as for the consistency of the $N$-soliton solution, it is
difficult to prove both of equation \eref{QR_ei} from the restrictions 
on the scattering
data \eref{z_z}--\eref{C_C}. In the continuous theory \cite{Tsuchida1}, 
\eref{QR_ei} is 
easily proved at least for $2 \times 2$ matrices $Q^{(2)}$ and
$R^{(2)}$. However, even in this case, it is not so easy to prove both 
of \eref{QR_ei} in the discrete theory.

Fourthly, due to the relation \eref{A_res}, we always have to consider 
pairs of poles as scattering data in the ISM. 
The constraint gives the novel structure of solutions, 
`two carrier waves in one envelope soliton'. 
This situation is closely related to the high internal
symmetries of $Q_n^{(m)}$ and $R_n^{(m)}$ which may give a rich variety
of dynamical behaviors in the 
multi-field soliton systems. As an influence of this fact, we need to assume 
additional conditions \eref{C_bar_C} to exclude breather-type solutions.

The sd-CNLS equations can be cast into alternative expressions by
some transformations. We set a pair of variables $q_n^{(j)}$ and 
$r_n^{(j)}$ to be constant in \eref{CNLS1}, for instance, 
$q_n^{(m)} r_n^{(m)} =1$ and consider a transformation of variables,
\[
\eql{
q_n^{(j)}\e^{2 \i t} \to q_n^{(j)}
\\
r_n^{(j)}\e^{-2 \i t} \to r_n^{(j)} 
}
\hspace{5mm}
j=1,2, \ldots, m-1 .
\]
Then, we obtain a simplified deformation of \eref{CNLS1},
\[
\eql{
\i \frac{\partial q^{(j)}_n}{\partial t}
 = \sum_{k=1}^{m-1} q^{(k)}_n r^{(k)}_n \cdot (q^{(j)}_{n+1}+q^{(j)}_{n-1})
\\
\i \frac{\partial r^{(j)}_n}{\partial t}
 = -\sum_{k=1}^{m-1} r^{(k)}_n q^{(k)}_n \cdot (r^{(j)}_{n+1}+r^{(j)}_{n-1})
}
\hspace{5mm}
j=1, 2, \ldots, m-1 .
\]

To consider a multi-field extension of the
 full-discrete NLS equation \cite{AS,AC,Suris} by our method, 
we need to modify the $L_n$-matrix \eref{U_form} appropriately. 
The details of the analysis will be reported in a separate paper.

\ack
One of the authors (TT) would like to express his sincere thanks to 
K Sogo, Y Ohta, 
M Hisakado, M Shiroishi and S Murakami for valuable discussions and
 comments. After completing the inverse scattering method, 
TT was informed by Hisakado 
that the Lax formulation based on a matrix representation of
 the Clifford algebra 
was introduced by Eichenherr and Pohlmeyer \cite{Pohlmeyer}. 
The authors are grateful to Y Ohta, D Trubatch and
 M J Ablowitz for notifying the authors of
 their results prior to publication. 
In order to check the integrals of motion
 of the sd-CNLS equations, the authors were partly helped by a beneficial 
{\it Mathematica} algorithm coded by \"{U} G\"{o}kta\c{s}, W 
Hereman and G Erdmann \cite{Goktas}. 

\appendix
\section*{Appendix. Trace formulae}
\setcounter{section}{1}
In this appendix, we show interrelations between the generator of the 
conserved densities $\tr \{ \log (I + Q_n \Gamma_n /\tz) \}$ 
for \eref{mNLS1} in section \ref{Conservation laws} and the scattering
data in section 3. 

Let us recall that we have transformed the scattering problem 
\eref{scattering0} 
into the scattering problem \eref{scattering_problem2} by the gauge 
transformation \eref{gauge}. 
%
For the moment, we do not impose restrictions 
such as \eref{eq_ref2} and \eref{R_Q} (or equation \eref{eq_ref91}) 
on square matrices $Q_n$ and $R_n$ 
and consider the sd-matrix NLS equation \eref{mNLS1}. 
The conditions 
\eref{F_H} are assumed. 
We supplement some definitions and relations. First, 
we define the inverse of \eref{ref111} by
\bea
\bar{\psi}_n (z) = \phi_n (z) {\cal A}(z) + 
                 \bar{\phi}_n (z) {\cal B}(z)
\nn \\
\psi_n (z) = \phi_n (z) \bar{\cal B}(z) -
                 \bar{\phi}_n (z) \bar{\cal A}(z).
\nn
\eea
Secondly, the generator of the conserved densities is 
invariant under the gauge transformation \eref{gauge}:
\bea
\fl &&
  \tr \log \Bigl( I + \frac{1}{z} \ti{Q}_n 
        \Phi_{2 \, n} \Phi_{1 \, n}^{-1} \Bigr)
\nn \\
\fl &=& \tr \log \Bigl\{
        I + \frac{1}{\tz \e^{\i \frac{\pi}{4}}} \cdot \e^{\i \frac{\pi}{2}n}
        (F_1)^n \e^{-\i H_1 t} Q_n \e^{\i H_2 t} (F_2)^n 
        \cdot \e^{-\i \frac{\pi}{4}(n-1)} (F_2)^n \e^{-\i H_2 t} 
        \Psi_{2\, n} \cdot \Psi_{1\, n}^{-1} \e^{\i H_1 t} (F_1)^n 
        \e^{-\i \frac{\pi}{4} n}  \Bigr\}
\nn \\
\fl &=& \tr \log \Bigl( I + \frac{1}{\tz}Q_n \Psi_{2\, n}\Psi_{1\, n}^{-1} 
        \Bigr) .
\eea
Thirdly, the asymptotic behaviors of the Jost functions $\phi_n$ and 
$\bar{\psi}_n$ are given by
 \bea
\fl
\hspace{30mm}
 \phi_n \equiv
 \left[
 \begin{array}{c}
   \phi_{1\, n}  \\
   \phi_{2\, n}  \\
 \end{array}
 \right]
 &\sim&
 \left[
 \begin{array}{c}
   I  \\
   O  \\
 \end{array}
 \right]
 z^n
 \hspace{14.8mm}
 {\rm as}~~~ n \rightarrow -\infty
\nn \\
\fl
\hspace{30mm}
\hspace{15mm}
&\sim& \left[
 \begin{array}{c}
   A(z) z^n    \\
   B(z) z^{-n} \\
 \end{array}
 \right]
 \hspace{7mm}
 {\rm as}~~~ n \rightarrow +\infty 
\nn
\eea
 \bea
\fl
\hspace{30mm}
 \bar{\psi}_n \equiv
 \left[
 \begin{array}{c}
   \bar{\psi}_{1\, n}  \\
   \bar{\psi}_{2\, n}  \\
 \end{array}
 \right]
 &\sim&
 \left[
 \begin{array}{c}
   {\cal A} (z) z^n    \\
   - {\cal B} (z) z^{-n} \\
 \end{array}
 \right]
 \hspace{7.2mm}
 {\rm as}~~~ n \rightarrow -\infty
\nn \\
\fl
\hspace{30mm}
\hspace{15mm}
&\sim& \left[
 \begin{array}{c}
   I  \\
   O  \\
 \end{array}
 \right]
 z^n
 \hspace{14.5mm}
 {\rm as}~~~ n \rightarrow +\infty.
\nn
\eea
Further, we can prove 
%
that $\phi_{2\, n} \phi_{1\, n}^{-1}$ is 
a polynomial in $1/z$ and $\bar{\psi}_{2\, n} \bar{\psi}_{1\, n}^{-1}$ 
is a polynomial in $z$. Therefore, we can replace 
$\phi_{2\, n} \phi_{1\, n}^{-1}$ and 
$\bar{\psi}_{2\, n} \bar{\psi}_{1\, n}^{-1}$ by $\Gamma_n^{(-)}$ and 
$\Gamma_n^{(+)}$ in section \ref{Conservation laws} respectively, 
except for the difference in gauge (see equation \eref{gauge}). 
It is important that the ratios of two components, 
$\phi_{2\, n} \phi_{1\, n}^{-1}$ and 
$\bar{\psi}_{2\, n} \bar{\psi}_{1\, n}^{-1}$, are invariant when 
we consider the time-dependent Jost functions 
$\phi_n^{(t)} \equiv \phi_n \e^{z^2 t}$ and 
$\bar{\psi}_n^{(t)} \equiv \bar{\psi}_n \e^{z^2 t}$. 

Now, we can relate the scattering data with the generator of the conserved
densities. The determinant of $A(z)$, $\det A(z)$, is expressed as
\bea
\fl && \log \det A(z) =  \tr \log A(z) 
\nn \\
\fl &=&  \tr \sum_{n=-\infty}^\infty \bigl[\log 
        (\phi_{1\, n+1} z^{-n-1}) 
        -\log (\phi_{1\, n} z^{-n}) \bigr] 
\nn \\
\fl &=& \tr \sum_{n=-\infty}^\infty \log
        \bigl[ \phi_{1\, n+1} \phi_{1\, n}^{-1} z^{-1} \bigr] 
\nn \\
\fl &=& \tr \sum_{n=-\infty}^\infty \log
        \Bigl[ I + \frac{1}{z} \ti{Q}_n \phi_{2\, n}
         \phi_{1\, n}^{-1} \Bigr] 
\nn \\
\fl &=& \tr \sum_{n=-\infty}^\infty \log
        \Bigl[ I + \frac{1}{\tz} Q_n \Gamma_n ^{(-)}
        \Bigr] 
\nn \\
\fl &=& \tr \sum_{n=-\infty}^\infty \Bigl[
\frac{1}{\tz^2} Q_n F_2 R_{n-1} F_1
+ \frac{1}{\tz^4}\{Q_n R_{n-2}-
 Q_n F_2 R_{n-1} Q_{n-1} F_2 R_{n-2} 
- \hf (Q_n F_2 R_{n-1} F_1)^2 \} + \cdots \Bigr].
\nn 
\\ \fl 
\label{trace1}
\eea
Similarly, $\det {\cal A}(z)$ is rewritten as
\bea
\fl && \log \det {\cal A} (z) 
\nn \\
\fl &=&  -\tr \sum_{n=-\infty}^\infty \bigl[\log 
        (\bar{\psi}_{1\, n+1} z^{-n-1}) 
        -\log (\bar{\psi}_{1\, n} z^{-n}) \bigr] 
\nn \\
\fl &=& -\tr \sum_{n=-\infty}^\infty \log
        \Bigl[ I + \frac{1}{z} \ti{Q}_n \bar{\psi}_{2\, n}
        \bar{\psi}_{1\, n}^{-1} \Bigr] 
\nn \\
\fl &=& -\tr \sum_{n=-\infty}^\infty \log
        \Bigl[ I + \frac{1}{\tz} Q_n \Gamma_n ^{(+)}
        \Bigr] 
\nn \\
\fl &=& \tr \sum_{n=-\infty}^\infty \Bigl[
        - \log (I-Q_n R_n) 
        + \tz^2 Q_n F_2 R_{n+1} F_1
\nn \\ 
\fl &&  \hspace{14mm}
 + \tz^4 \{Q_n R_{n+2}
 - Q_n R_{n+2} F_1 Q_{n+1} R_{n+1} F_1  
 - \hf (Q_n F_2 R_{n+1} F_1)^2 \} + \cdots \Bigr].
\label{trace2}
\eea
Here the time independence of ${\cal A} (z)$, 
${\cal A}_t (z) =O$, is proved in the same manner as in section 
\ref{Time-dep}. It is now clear how the scattering data are expressed 
in terms of the integrals of motion for the sd-matrix NLS equation. 
However, it should be stressed that the above expansions do not yield 
local conservation laws. The method presented in section 
\ref{Conservation laws} is useful because it gives not only the densities 
but also the corresponding fluxes.

Conversely, the integrals of
motion can be expressed in terms of the scattering data. For
simplicity, we assume that

i) $\ti{Q}_n$ and $\ti{R}_n$ are expressed as (\ref{Q_R_const}). 
Thus, proposition 3.$\,$4.$\,$1 holds.

ii) $\det A(z)$ and $\det \bar{A} (z)$ have $4N$ simple zeros 
outside and inside the unit circle $C$, 

$\hspace{3.5mm}$ respectively. None of them lies on the unit circle $C$.

iii) $\det A(z)$ and $\det \bar{A} (z)$ approach $1$ 
rapidly as $|z| \to \infty$ and $z \to 0$, respectively.
\\
Then, we can derive the following expansion 
for the sd-CNLS equations \eref{CNLS4}:
%
\bea
\fl && \log \det  A(z)
\nn \\
\fl &=&  \sum_{n=1}^\infty 
\frac{1}{z^{n}} \Bigl[
\frac{1}{n} \sum_{j=1}^{4N} \Bigl\{ 
\Bigl( \frac{1}{z_j^\ast}\Bigr)^{n} - z_j^{n} \Bigr\} 
 + \frac{1}{2 \pi \i} \oint_{C} w^{n-1} \log \det (A(w) \bar{A}(w)) \d w
\Bigr]
\nn \\
\fl &=&  \sum_{k=1}^\infty 
\frac{1}{z^{2k}} \Bigl[
\frac{1}{k} \sum_{j=1}^{2N} \Bigl\{ 
\Bigl( \frac{1}{z_j^\ast}\Bigr)^{2k} - z_j^{2k} \Bigr\} 
+ \frac{1}{\pi \i} \int_{C_R} w^{2k-1} \log \det (A(w) \bar{A}(w)) \d w
\Bigr]
\nn \\
\fl &=&  \sum_{k=1}^\infty 
\frac{1}{z^{2k}} \Bigl[
\frac{1}{k} \sum_{j=1}^{N} \Bigl\{ 
\Bigl( \frac{1}{z_j}\Bigr)^{2k} + \Bigr(\frac{1}{z_j^\ast}\Bigr)^{2k} 
        - z_j^{2k} - z_j^{\ast \, 2k} \Bigr\} 
 + \frac{1}{\pi \i} \int_{C_{UR}} 
 ( w^{2k-1} +w^{-2k-1} ) \log | \det A(w) |^2  \d w
\Bigr] .
\nn \\
\fl
\label{trace3}
\eea
%
%
%
%
The coefficients of $1/z^{2k}\; (k=1, 2, \ldots)$ give an infinite 
number of the integrals of motion, which we call the 
{\it trace formulae}. Here, $C_R$ and $C_{UR}$ 
denote the right-half portion and the 
upper-right portion of the unit circle $C$, respectively, as is 
mentioned in section 3. It is recalled that $\tz$ in 
\eref{trace1} and $z$ in \eref{trace3} have the difference $\pi/4$ in their
phases (see equation \eref{QR_trans}). 
The determinant of $\bar{A}(z)$ is related
to the determinant of $A(z)$ by \eref{det_A2}. It can also be shown
that $\det {\cal A}(z)$ and $\det \bar{A} (z)$ are
 connected by
\[
\log \det {\cal A}(z) = - \sum_{n=-\infty}^\infty 
\log \det (I-Q_n R_n) + \log \det \bar{A}(z).
\]
Thus, we can directly obtain expansions of $\log \det \bar{A} (z)$ 
and $\log \det {\cal A}(z)$ with respect to $z$ from \eref{trace3}.

A derivation of \eref{trace3} is omitted because it is
analogous to that in the continuous theory \cite{AS,ZM,FN}. 
Related results were also obtained by Kodama \cite{Kodama}.


\section*{References}
\begin{thebibliography}{99}

\bibitem{ZS1}
Zakharov V E and Shabat A B 1972 \SPJ {\bf 34} 62

\bibitem{ZS2}
Zakharov V E and Shabat A B 1973 \SPJ {\bf 37} 823

\bibitem{Wadati0}
Wadati M 1972 \JPSJ {\bf 32} 1681

\bibitem{AKNS}
Ablowitz M J, Kaup D J, Newell A C and Segur H
 1973 \PRL {\bf 31} 125

\bibitem{AS}
Ablowitz M J and Segur H 1981 {\it Solitons and the
Inverse Scattering Transform} (Philadelphia, PA: SIAM)

\bibitem{AC} Ablowitz M J and Clarkson P A 1991
 {\it Solitons, Nonlinear Evolution Equations and Inverse Scattering}
 (Cambridge: Cambridge University Press)






\bibitem{Berk}
Berkhoer A L and Zakharov V E 1970 \SPJ {\bf 31} 486

\bibitem{Manakov}
Manakov S V 1974 \SPJ {\bf 38} 248

\bibitem{Shulman}
Zakharov V E and Schulman E I 1982 {\it Physica} D {\bf 4} 270

\bibitem{Ohta2}
Ohta Y 1989 {\it RIMS k\^{o}ky\^{u}roku} {\bf 684} 1 (in Japanese)

\bibitem{RL}
Radhakrishnan R and Lakshmanan M 1995 \JPA {\bf 28} 2683

\bibitem{Menyuk}
Menyuk C R 1987 {\it IEEE J. Quantum Electron.} {\bf 23} 174

\bibitem{Kivshar}
Kivshar Y S and Turitsyn S K 1993 {\it Opt. Lett.} {\bf 18} 337

\bibitem{Yang}
Yang J 1997 {\it Physica} D {\bf 108} 92

\bibitem{Inoue}
Inoue Y 1976 {\it J. Plasma Phys.} {\bf 16} 439

\bibitem{Hisakado1}
Wadati M, Iizuka T and Hisakado M 1992 \JPSJ {\bf 61} 2241

\bibitem{Hisakado2}
Hisakado M, Iizuka T and Wadati M 1994 \JPSJ {\bf 63} 2887

\bibitem{Karlsson}
Karlsson M, Kaup D J and Malomed B A 1996 \PR E {\bf 54} 5802

\bibitem{RLH}
Radhakrishnan R, Lakshmanan M and Hietarinta J 1997 \PR E {\bf 56}
2213

\bibitem{Tsuchida1}
Tsuchida T and Wadati M 1998 \JPSJ {\bf 67} 1175

\bibitem{Svinolupov}
Svinolupov S I 1992 {\it Commun.\ Math.\ Phys.} {\bf 143} 559

\bibitem{Iwao}
Iwao M and Hirota R 1997 \JPSJ {\bf 66} 577

\bibitem{Hirota3}
Hirota R 1973 \JMP {\bf 14} 805



\bibitem{Case}
Case K M and Kac M 1973 \JMP {\bf 14} 594

\bibitem{Flaschka}
Flaschka H 1974 \PTP {\bf 51} 703

\bibitem{Manakov2}
Manakov S V 1975 \SPJ {\bf 40} 269

\bibitem{AL1}
Ablowitz M J and Ladik J F 1975 \JMP {\bf 16} 598

\bibitem{AL2}
Ablowitz M J and Ladik J F 1976 \JMP {\bf 17} 1011

\bibitem{Levi}
Levi D, Ragnisco O and Bruschi M 1980 {\it Nuovo Cimento} A {\bf 58} 56

\bibitem{Common}
Common A K 1992 \IP {\bf 8} 393

\bibitem{Veks}
Vekslerchik V E and Konotop V V 1992 \IP {\bf 8} 889

\bibitem{Ahmad}
Ahmad S and Chowdhury A R 1987 \JPA {\bf 20} 293

\bibitem{FD}
Christiansen P L, Eilbeck J C and Parmentier R D (ed) 1993 
{\it Future Directions of Nonlinear Dynamics in Physical and 
Biological Systems} (New York: Plenum)

\bibitem{Ross}
Ross K A and Thompson C J 1986 {\it Physica} A {\bf 135} 551

\bibitem{Bishop}
Cai D, Bishop A R and Gr\o nbech-Jensen N 1995 \PR E {\bf 52} 5784

\bibitem{Li}
Li Y 1992 \PL A {\bf 163} 181

\bibitem{AH} 
Ablowitz M J and Herbst B M 1990 {\it SIAM
J. Appl. Math.} {\bf 50} 339

\bibitem{Umeki1}
Umeki M 1996 {\it RIMS k\^{o}ky\^{u}roku} {\bf 974} 21

\bibitem{Umeki2}
Umeki M 1998 Homoclinic Structures in a Discrete Integrable System 
{\it Preprint}

\bibitem{Doliwa}
Doliwa A and Santini P M 1995 \JMP {\bf 36} 1259

\bibitem{Hisakado5}
Hisakado M and Wadati M 1996 \JPSJ {\bf 65} 389

\bibitem{Kulish}
Kulish P P 1981 \LMP {\bf 5} 191

\bibitem{FT} 
Faddeev L D and Takhtajan L A 1987
 {\it Hamiltonian Methods in the Theory of Solitons} (Berlin:
Springer)

\bibitem{Suris}
Suris Y B 1997 \IP {\bf 13} 1121

\bibitem{Hisakado4}
Hisakado M 1997 \JPSJ {\bf 66} 1939

\bibitem{Ohta}
Ohta Y 1997 Pfaffian Solution for Coupled Discrete Nonlinear 
Schr\"{o}dinger Equation {\it Chaos, Solitons and Fractals} at press

\bibitem{Merola}
Merola I, Ragnisco O and Gui-Zhang T 1994 \IP {\bf 10} 1315.

\bibitem{Tsuchida2}
Tsuchida T, Ujino H and Wadati M 1998 \JMP {\bf 39} 4785

\bibitem{Hirota2}
Hirota R 1997 \JPSJ {\bf 66} 2530

\bibitem{Trubatch}
Ablowitz M J, Ohta Y and Trubatch A D 1998 {\it Preprint} 
solv-int 9810014

\bibitem{Trubatch2}
Ablowitz M J, Ohta Y and Trubatch A D 1998 
On Integrability and Chaos in Discrete Systems 
{\it Chaos, Solitons and Fractals} at press


\bibitem{Wadati3}
Wadati M and Kamijo T 1974 \PTP {\bf 52} 397

\bibitem{Fordy}
Fordy A P and Kulish P P 1983 \CMP {\bf 89} 427

\bibitem{Pohlmeyer}
Eichenherr H and Pohlmeyer K 1979 \PL B {\bf 89} 76

\bibitem{Goktas}
G\"{o}kta\c{s} \"{U}, Hereman W and Erdmann G 1997 \PL A {\bf 236} 30

\bibitem{ZM}
Zakharov V E and Manakov S V 1974 Theor. Math. Phys. {\bf 19} 551

\bibitem{FN}
Flaschka H and Newell A C 1975 {\it Dynamical Systems, Theory and
Applications} ({\it Lecture Notes in Physics} vol 38) ed J Moser
(New York: Springer) p 355

\bibitem{Kodama}
Kodama Y 1975 \PTP {\bf 54} 669

\end{thebibliography}
\end{document}